\documentclass[12pt]{article}
\pdfoutput=1

\usepackage{multirow}
\usepackage[pdftex]{graphicx}
\usepackage{amssymb}
\usepackage{amsmath}
\usepackage{cite}
\usepackage{xspace}
\usepackage{slashed}
\usepackage{bbold}
\usepackage{setspace}
\usepackage{longtable}
\usepackage{booktabs}
\usepackage{array}
\usepackage{xcolor}
\definecolor{tit}{rgb}{0.1,0.2,0.4}
\definecolor{pol}{rgb}{0,0.4,0}
\definecolor{verde}{cmyk}{0.92,0,0.59,0.25}

\renewcommand{\arraystretch}{1.2}

\newcommand{\eq}[1]{\begin{equation} #1 \end{equation}}

\newcommand{\eqa}[1]{\begin{eqnarray} #1 \end{eqnarray}}

\newcommand{\av}[1]{\langle #1 \rangle}

\newcommand{\GeV}{\,{\rm GeV}}

\newcommand{\cO}{\mathcal{O}}
\newcommand{\cL}{\mathcal{L}}
\newcommand{\Eq}[1]{Eq.~(\ref{#1})}

\newcommand{\Sec}[1]{Section~\ref{#1}}
\newcommand{\App}[1]{Appendix~\ref{#1}}
\newcommand{\Reff}[1]{Ref.~\cite{#1}}

\newcommand{\red}[1]{{\color{red} #1}}

\begin{document}

$\ $
\vspace{1.5cm}

\begin{center}
\fontsize{20}{24}\selectfont
%\Large
\bf 
On the two-loop penguin contributions\\
to the Anomalous Dimensions\\
of four-quark operators
\end{center}

\vspace{2mm}

\begin{center}
{\rm
Pol Morell\,
and\, 
Javier Virto
}\\[5mm]
{\it\small
Departament de Física Quàntica i Astrofísica, Universitat de Barcelona,\\
Martí i Franquès 1, E08028 Barcelona, Catalonia\\[2mm]
Institut de Ciències del Cosmos (ICCUB), Universitat de Barcelona,\\
Martí i Franquès 1, E08028 Barcelona, Catalonia
}
\end{center}

\vspace{1mm}
\begin{abstract}\noindent
\vspace{-5mm}

We revisit the Next-to-Leading Order (two-loop) contributions to the Anomalous Dimensions of $\Delta F = 1$ four-quark operators in QCD.
We devise a test for anomalous dimensions, that we regard as of general interest, and by means of which we detect a problem in the results available in the literature. Deconstructing the steps leading to the available result, we identify the source of the problem, which is related to the operator known as~$Q_{11}$. We show how to fix the problem and provide the corrected anomalous dimensions. With the insight of our findings, we propose an alternative approach to the one used in the literature which does not suffer from the identified disease, and which confirms our corrected results. We assess the numerical impact of our corrections, which happens to be in the ballpark of $5\%$ in certain entries of the evolution matrix. Our results are important for the correct resummation of Next-to-Leading Logarithms in analyses of physics beyond the Standard Model in $\Delta F = 1$ processes, such as the decays of Kaons and $B$-mesons.

\end{abstract}

\newpage

\setcounter{tocdepth}{2}
\tableofcontents

\allowdisplaybreaks

%%%%%%%%%%%%%%%%%%%%%%%%%%%%%%%%%%%%%%%%%%%%%%%
\section{Introduction}

Particle-physics processes at energies significantly lower than the Electroweak (EW) scale --such as weak decays of hadrons-- are described by an Effective Field Theory (EFT) where EW-scale Standard Model (SM) particles as well as potential heavy Beyond-the-SM (BSM) fields are integrated out. The EFT description is very convenient in order to resum large logarithms that arise from the large hierarchy between the EW scale and the energy of the process (e.g. $m_B$ for a $B$ decay). Such logarithms can spoil the convergence of perturbation theory, particularly in QCD at energies below $\sim 5\GeV$, where the strong coupling is not small. The resummation of these logarithms is done by solving the Renormalization Group Equations (RGEs), in terms of the Anomalous Dimensions of the effective operators~\cite{Buchalla:1995vs,Aebischer:2017gaw}. Over the last three decades, significant efforts have been devoted to the calculation of EFT anomalous dimensions at two, three and even four loops in QCD.

Two-loop anomalous dimensions for $\Delta F=1$ four-quark operators of the type $(\bar s d)(\bar q q)$ were first calculated by Buras, Jamin, Lautenbacher and Weisz (BJLW) in the 1990's~\cite{Buras:1989xd, Buras:1992tc}. The calculation focused exclusively on the SM operator basis, and was performed both in the Naive Dimensional Regularization (NDR) and t'Hooft-Veltman (HV) schemes.
In order to avoid the usual problems involving traces with $\gamma_5$ in the NDR scheme, BJLW devised a method (hereon the ``BJLW method'') that only requires the calculation of penguin diagrams without closed fermion loops, where no ambiguous Dirac traces appear. The full set of anomalous dimensions can then be reconstructed from this reduced subset of diagrams.
This calculation was checked and confirmed in several subsequent papers using different operator bases and approaches~\cite{Ciuchini:1993vr, Ciuchini:1993fk, Bobeth:2003at, Gorbahn:2004my,Huber:2005ig}, some of them addressing the issues with $\gamma_5$ by using the well-known CMM scheme~\cite{Chetyrkin:1997gb}, where Dirac traces in the NDR scheme never contain a $\gamma_5$.
The results thus obtained
can be compared to one another by performing a change of basis properly at next-to-leading order~(NLO), accounting for the appropriate scheme dependence, including evanescent terms.

In the seminal paper by Buras, Misiak and Urban (BMU)~\cite{Buras:2000if}, this set of anomalous dimensions was extended to the full basis Beyond the Standard Model (BSM). 
This basis includes three additional operators that complete the set involving penguin diagrams\,\footnote{
Following~\Reff{Buras:2000if} we will focus on the case of $\bar s\to \bar d$ transitions, as a proxy to all other $\Delta F=1$ sectors.
}
:
\eqa{
Q_{11} &=&
(\bar{s}^{\alpha} \gamma^{\mu}P_L d^{\alpha}) (\bar{s}^{\beta} \gamma_{\mu}P_L s^{\beta}) + 
(\bar{s}^{\alpha} \gamma^{\mu}P_L d^{\alpha}) (\bar{d}^{\beta} \gamma_{\mu}P_L d^{\beta})
\ ,
\\
Q_{12} &=&
(\bar{s}^{\alpha} \gamma^{\mu}P_L d^{\beta}) (\bar{s}^{\beta} \gamma_{\mu}P_R s^{\alpha}) + 
(\bar{s}^{\alpha} \gamma^{\mu}P_L d^{\beta}) (\bar{d}^{\beta} \gamma_{\mu}P_R d^{\alpha})
\ ,
\\
Q_{13} &=&
(\bar{s}^{\alpha} \gamma^{\mu}P_L d^{\alpha}) (\bar{s}^{\beta} \gamma_{\mu}P_R s^{\beta}) + 
(\bar{s}^{\alpha} \gamma^{\mu}P_L d^{\alpha}) (\bar{d}^{\beta} \gamma_{\mu}P_R d^{\beta})
\ ,
}
as well as the three corresponding operators with opposite chirality. 
The penguin contributions to the anomalous dimensions of these operators were obtained by BMU from the SM subset computed by BJLW, in a procedure analogous to the BJLW method.
In this way, BMU provided the complete NLO (two-loop) QCD Anomalous Dimension Matrix (ADM) in the general BSM case. These results, to the best of our knowledge, have never been confirmed independently.

However, as we shall discuss in the following, there is a class of tests that can be carried out in any ADM calculation, based on the fact that anomalous dimensions satisfy a specific form of flavor symmetry. In the case at hand, this flavor symmetry ensures that under a transformation changing quark flavors $u \leftrightarrow b$, the ADM must remain the same,
\eq{
\label{BMU=BMU'pre}
\hat \gamma_\text{BMU} = \hat \gamma_{\text{BMU}^\prime}\ ,
}
where BMU${}^\prime$ is an operator basis obtained from the operator basis in BMU by performing the field replacements $u\leftrightarrow b$ everywhere.
This condition is non-trivial, and obtaining $\hat \gamma_{\text{BMU}^\prime}$ from $\hat \gamma_\text{BMU}$ requires a complete knowledge of the renormalization scheme in which $\hat \gamma_\text{BMU}$ is given.
The ADM for the SM sector given in BJLW satisfies this condition exactly,
but the ADM including BSM operators in BMU, assuming our interpretation of the scheme used therein, explicitly violates~\Eq{BMU=BMU'pre}.

The purpose of this paper is to raise, clarify, and resolve this issue, and to provide the correct two-loop ADM for the $\Delta F=1$ sector. 
We will also provide some insights that may be useful in checking and manipulating anomalous dimension matrices.
We shall see that the particular way in which the BJLW method is extended in~\Reff{Buras:2000if} is in fact not valid, but that it can be modified minimally by introducing, in an intermediate step, a symmetrized operator $Q_{11}^+$, leading to an ADM that satisfies the flavor symmetry condition in~\Eq{BMU=BMU'pre}.

This letter is organized as follows.
We begin in~\Sec{sec:set-up} reviewing the necessary formalism regarding the NLO renormalization of the EFT.
In~\Sec{sec:problem} we present the problem: why the ADM presented in~\Reff{Buras:2000if} presents an inconsistency.
In~\Sec{sec:diagnosis} we diagnose the problem, showing that it is related to the anomalous dimension of the operator $Q_{11}$ and more precisely to the relation in~\Eq{gammaQ11} below.
The corrected entries of the ADM are presented in~\Sec{sec:solution},
where we also show that our ADM satisfies the flavor symmetry condition, thus solving the problem raised.
Based on the insight gained, in~\Sec{sec:proposal} we present a proposal for a correct alternative to the approach in~\Reff{Buras:2000if}, and show that this alternative expression does indeed provide the correct result for the NLO ADM.
In order to gauge the numerical importance of the corrected anomalous dimensions, in~\Sec{sec:numerics} we perform a simple numerical analysis. Finally, we conclude in~\Sec{sec:conclusion} with a summary of our results.

%%%%%%%%%%%%%%%%%%%%%%%%%%%%%%%%%%%%%%%%%%%%%%%
\section{Renormalization of the Effective Theory}
\label{sec:set-up}

The renormalized EFT Lagrangian is given by
\eq{
\cL_\text{EFT}
= \cL_\text{QCD} + \sum_{i,j} C_i Z_{ij} Z_q^2 \cO_j \ ,
}
where the (renormalized) operators $\cO = \{ Q , E \}$ include physical ($Q_i$) as well as evanescent ($E_i$) operators, the latter needed for renormalization in $d=4-2\epsilon$ dimensions.
The operators relevant for $\Delta F=1$ transitions in the so-called ``BMU basis'' of~\Reff{Buras:2000if} are given in~\App{app:OPBasis}. The $C_i$ are renormalized Wilson coefficients, and $Z_{ij}$ is the renormalization constant matrix, which takes care of the renormalization of the Wilson coefficients and it is responsible for operator mixing. The renormalization factor $Z_q$ takes care of quark wave-function renormalization of the four-quark operators (one factor of $Z_q^{1/2}$ for each field).

The renormalized Wilson coefficients depend on the renormalization scale as
\eq{
\label{RGE}
\frac{dC_i}{d\log\mu} = \gamma_{ji}\,C_j = (\hat\alpha_s \gamma^{(0)}_{ji} + \hat\alpha_s^2 \gamma^{(1)}_{ji} + \cdots)\,C_j
\ ,
}
where $\hat\gamma$ (with components $\gamma_{ij}$) is the Anomalous Dimension Matrix (ADM), and $\hat\gamma^{(i)}$ are the constant coefficients in its expansion in powers of $\hat\alpha_s\equiv g_s^2/(4\pi)^2$. In terms of the renormalization matrix,
\eq{
\hat\gamma = \hat Z \frac{d \hat Z^{-1}}{d\log\mu}
\ ,
}
with $\hat Z$ depending on the renormalization scale through its expansion in $\hat\alpha_s(\mu)$,
\eq{
\label{eq:Z Expansion}
\hat Z = 1 + \sum_{\ell=1}^\infty \sum_{m=0}^\ell \frac{\hat \alpha_s^{\ell}}{\epsilon^m} \,\hat Z^{(\ell,m)}\ .
}
In the $\overline{\text{MS}}$ scheme, $Z_{ij}^{(\ell,0)}=0$ whenever $i$ refers to a physical operator or $j$ refers to an evanescent one.
With this notation at hand, one finds (see e.g.~\Reff{Gorbahn:2004my})
\eqa{
\hat\gamma^{(0)} &=& 2\hat Z^{(1,1)}\ , 
\label{gamma0Z}\\
\hat\gamma^{(1)} &=& 4\hat Z^{(2,1)} - 2\hat Z^{(1,1)}\hat Z^{(1,0)}\ ,
\label{gamma1Z}
}
and so on. The renormalization constants can be calculated in the $\overline{\text{MS}}$ scheme in terms of the (amputated) renormalized\footnote{By renormalized amplitudes or diagrams we mean with respect to $\mathcal{L}_{\rm QCD}$, i.e. including both the bare diagrams and the counterterms from the $\text{dim} \leq 4$ Lagrangian.} matrix elements of the operators $Q_i$. At any loop order, we write
\eqa{
\av{Q_i} &=&
\sum_{\ell=0}^\infty
\tilde\mu^{2\ell\epsilon}\, 
\hat \alpha_s^\ell \ \av{Q_i}^{(\ell)}\ ,
\\
\av{Q_i}^{(\ell)} &=& 
\sum^{\ell}_{k = 0} \frac{1}{\epsilon^{k}}
\left[ a^{(\ell,k)}_{Q_iQ_j} \langle Q_j\rangle^{(0)}
+ a^{(\ell,k)}_{Q_iE_j} \langle E_j \rangle^{(0)} \right] 
\ ,
}
where $\tilde\mu$ is the $\overline{\text{MS}}$ scale.
The running of $\alpha_s$ is given at the leading order by the 
%renormalization factor of $g_s$:
%$Z_g = 1 - \frac1\epsilon\hat\alpha_s \beta_0 + \cO(\hat\alpha_s^2)$, with 
QCD beta function $\beta_0 = \frac{11}3 N_c - \frac23 f$ (see e.g.~\Reff{Buchalla:1995vs}), where $f$ is the number of active quark flavors.
The coefficients $a_{Q_i \cO_j}^{(\ell,k)}$ arise from the $1/\epsilon^k$ poles of the renormalized $\ell$-loop diagrams with insertion of operator $Q_i$.
The renormalization then leads to
\eqa{
\hat Z^{(1,1)} &=& -\hat a^{(1,1)} - 2Z_q^{(1,1)}\hat{\mathbb{1}}\ , \\
\hat Z^{(2,1)} &=& -\hat a^{(2,1)} + \hat a^{(1,1)}\cdot \hat a^{(1,0)} - \hat Z^{(1,0)}\cdot \hat a^{(1,1)}
- 2Z_q^{(2,1)}\hat{\mathbb{1}}\ ,
}
up to two loops. Here we have expanded $Z_q$ as in \Eq{eq:Z Expansion}, and its coefficients read
\eq{
Z_q^{(1,1)} = -C_F \;, 
\qquad Z_q^{(2,1)} = 
C_F\bigg[
\,\frac{3}{4}C_F - \frac{17}{4}N_c + \frac{1}{2}f\,
\bigg] 
\;.
}
In addition, in the Buras-Weisz scheme for evanescent operators, we have that $Z^{(1,0)}_{ij}=-a^{(1,0)}_{ij}$ for $(i,j)$=(evanescent,physical), and zero otherwise.
Inserting these expressions into~Eqs.\,(\ref{gamma0Z})-(\ref{gamma1Z}) one finds
\eqa{
\gamma^{(0)}_{ij} &=&
-2 \hat{a}^{(1,1)}_{Q_iQ_j} - 4Z_q^{(1,1)} \delta_{Q_iQ_j}\ ,
\label{gamma0a}\\
\gamma^{(1)}_{ij} &=&
-4 \hat{a}^{(2,1)}_{Q_iQ_j}
+4 \hat{a}^{(1,1)}_{Q_iQ_k}\hat{a}^{(1,0)}_{Q_kQ_j}
+2 \hat{a}^{(1,1)}_{Q_iE_k}\hat{a}^{(1,0)}_{E_kQ_j} 
-8 Z_q^{(2,1)} \delta_{Q_iQ_j}
\ .
\label{gamma1a}
}
In~\Reff{Buras:2000if}, BMU give the full results for $\gamma^{(0)}$ and $\gamma^{(1)}$ for the full operator basis.
However, these results fail a simple consistency test, as we shall explain in the following section.

%%%%%%%%%%%%%%%%%%%%%%%%%%%%%%%%%%%%%%%%%%%%%%%
\section{The Problem: a flavor symmetry}
\label{sec:problem}

We are going to consider the ADM in two different bases. The first one is the BMU basis, as given in~\App{app:OPBasis}, while the second one is a modified version (BMU${}^\prime$) defined simply as
\eq{
Q_i^{(\text{BMU}^\prime)} = Q_i^{\text{(BMU)}}\Big|_{u \leftrightarrow b} \;.
}
In dimensional regularization, the ADMs can be calculated by setting to zero the quark masses, given that they depend exclusively on the UV structure of the theory. Thus, the difference between BMU and BMU${}^\prime$ is merely a `renaming' of up and bottom quark fields. Hence, the ADM should have the exact same explicit entries before and after the renaming:
\eq{
\hat{\gamma}_{\text{BMU}^\prime} = \hat{\gamma}_{\text{BMU}}\ .
}
This relation can be checked by explicitly performing a change of basis. Note that this change of basis is very non-trivial and involves Fierz-evanescent operators. Up to NLO~\cite{Buras:1991jm,Gorbahn:2004my,Chetyrkin:1997gb}, 
\eqa{
\hat{\gamma}^{(0)}_{\text{BMU}^\prime} 
&=&
\hat{R} \hat{\gamma}^{(0)}_\text{BMU} \hat{R}^{-1}\ ,
\\
\hat{\gamma}^{(1)}_{\text{BMU}^\prime}
&=& \hat{R} \hat{\gamma}^{(1)}_\text{BMU}\hat{R}^{-1}
- 2\beta_0 \Delta\hat{r} - \left[\Delta\hat{r}, \hat{\gamma}^{(0)}_{\text{BMU}^\prime}\right] \ .
}
The correct NLO ADM should satisfy the following condition,
\eq{
\label{ubsym}
\hat{\gamma}^{(1)}_{\text{BMU}} \quad \xrightarrow[\substack{\text{NLO Change} \\ \text{of Basis}}]{} \quad \hat{\gamma}^{(1)}_{\text{BMU}^\prime} \quad\equiv\quad \hat{\gamma}^{(1)}_{\text{BMU}} \ .
}
The details of the transformation involve calculating the tree-level transformation matrix $\hat{R}$ and the evanescent shift in the renormalization scheme, $\Delta\hat{r}$. 
For the latter we use the $\overline{\text{MS}}$-NDR scheme with the Buras-Weisz prescription~\cite{Buras:1989xd}, combined with the basis of evanescent operators given below in~\App{app:EVBasis}.
This basis of evanescent operators is equivalent\,\footnote{
Even though BMU do not give explicitly in~\Reff{Buras:2000if} the evanescent basis used for $\bar{s}d\bar{q}q$ operators, the fact that their current-current contributions to the ADM are taken directly (and explicitly) from sectors $\bar{s}d\bar{u}c$ and $\bar{s}d\bar{s}d$ ---for which they \emph{do} present the evanescent basis--- allows us to infer their scheme. See the discussion in~\Sec{sec:diagnosis} for further details.
} to the one used in~Refs.~\cite{Buras:1992tc, Buras:2000if},
and corresponds both to the use of \textit{Greek projections} and also to the choice $a_{\text{ev}},b_{\text{ev}},c_{\text{ev}}, ... = 1$ in~\Reff{Dekens:2019ept}. We also adopt this scheme in all our calculations throughout this work.

We focus on the sector of vector operators $\{Q_1,Q_2,...,Q_{18}\}$ which is where the problem arises.
The tree-level transformation matrix in this sector is given by
\eq{
\hat{R} = \left(
\setlength{\arraycolsep}{4.2pt}
\renewcommand{\arraystretch}{1.2}
\begin{array}{cccccccccccccccccc}
0 & 0 & \frac{2}{3} & 0 & 0 & 0 & 0 & 0 & -\frac{2}{3} & 0 & -1 & 0 & 0 & 0 & 0 & 0 & 0 & 0
\\
0 & 0 & 0 & \frac{2}{3} & 0 & 0 & 0 & 0 & 0 & -\frac{2}{3} & -1 & 0 & 0 & 0 & 0 & 0 & 0 & 0
\\
0 & 0 & 1 & 0 & 0 & 0 & 0 & 0 & 0 & 0 & 0 & 0 & 0 & 0 & 0 & 0 & 0 & 0 \\
0 & 0 & 0 & 1 & 0 & 0 & 0 & 0 & 0 & 0 & 0 & 0 & 0 & 0 & 0 & 0 & 0 & 0 \\
0 & 0 & 0 & 0 & 1 & 0 & 0 & 0 & 0 & 0 & 0 & 0 & 0 & 0 & 0 & 0 & 0 & 0 \\
0 & 0 & 0 & 0 & 0 & 1 & 0 & 0 & 0 & 0 & 0 & 0 & 0 & 0 & 0 & 0 & 0 & 0 \\
0 & 0 & 0 & 0 & \frac{3}{4} & 0 & -\frac{1}{2} & 0 & 0 & 0 & 0 & 0 & -\frac{3}{2} & 0 & 0 & 0 & 0 & -\frac{3}{4}
\\
0 & 0 & 0 & 0 & 0 & \frac{3}{4} & 0 & -\frac{1}{2} & 0 & 0 & 0 & -\frac{3}{2} & 0 & 0 & 0 & 0 & -\frac{3}{4} & 0
\\
-\frac{3}{2} & 0 & 1 & 0 & 0 & 0 & 0 & 0 & 0 & 0 & -\frac{3}{2} & 0 & 0 & 0 & 0 & 0 & 0 & 0
\\
0 & -\frac{3}{2} & 0 & 1 & 0 & 0 & 0 & 0 & 0 & 0 & -\frac{3}{2} & 0 & 0 & 0 & 0 & 0 & 0 & 0
\\
0 & 0 & 0 & 0 & 0 & 0 & 0 & 0 & 0 & 0 & 1 & 0 & 0 & 0 & 0 & 0 & 0 & 0
\\
0 & 0 & 0 & 0 & 0 & 0 & 0 & 0 & 0 & 0 & 0 & 1 & 0 & 0 & 0 & 0 & 0 & 0
\\
0 & 0 & 0 & 0 & 0 & 0 & 0 & 0 & 0 & 0 & 0 & 0 & 1 & 0 & 0 & 0 & 0 & 0
\\
0 & 0 & 0 & 0 & 0 & 0 & 0 & 0 & 0 & 0 & 0 & 0 & 0 & 1 & 0 & 0 & 0 & 0
\\
0 & 0 & 0 & 0 & 0 & 0 & 0 & 0 & 0 & 0 & 0 & 0 & 0 & 0 & 1 & 0 & 0 & 0
\\
0 & 0 & 0 & 0 & 0 & 0 & 0 & 0 & 0 & 0 & 0 & 0 & 0 & 0 & 0 & 1 & 0 & 0
\\
0 & 0 & 0 & 0 & 0 & 0 & 0 & 0 & 0 & 0 & 0 & 0 & 0 & 0 & 0 & 0 & -1 & 0
\\
0 & 0 & 0 & 0 & 0 & 0 & 0 & 0 & 0 & 0 & 0 & 0 & 0 & 0 & 0 & 0 & 0 & -1 
\\
\end{array}
\right) \ ,
}
while the matrix $\Delta \hat r$ only has two non-zero rows, with non-zero entries on the columns corresponding to the four QCD penguin operators,
\eq{
\setlength{\arraycolsep}{4.5pt}
\renewcommand{\arraystretch}{1.5}
\begin{array}{ccccccccccccc}
{[\Delta \hat r]}_{2\,j} & =
& \big( & 0 & 0 & \textstyle{\frac{1}{N_c}} & -1 & \textstyle{\frac{1}{N_c}} & -1 & 0 &\cdots & 0 & \big)\ ,
\\
{[\Delta \hat r ]}_{10\,j} & =
& \big( & 0 & 0 & \textstyle{\frac{3}{2N_c}} & \textstyle{-\frac32} & \textstyle{\frac{3}{2N_c}} & \textstyle{-\frac32} & 0 & \cdots & 0 & \big)\ .
\end{array}
}
For the LO ADM one finds indeed that $\gamma^{(0)}_\text{BMU} = \gamma^{(0)}_{\text{BMU}^\prime}$.
However, at NLO, implementing the change of basis starting with the original $\gamma^{(1)}_\text{BMU}$ in~\Reff{Buras:2000if} leads to a direct violation of~\Eq{ubsym}, as $\gamma^{(1)}_\text{BMU}$ and $\gamma^{(1)}_{\text{BMU}^\prime}$ are found to differ in the QCD penguin columns ($Q_{3},Q_{4},Q_{5},Q_{6}$) and rows first, second, ninth and tenth:
\eq{
\label{BMU-BMU'}
\gamma^{(1)}_\text{BMU}-\gamma^{(1)}_{\text{BMU}^\prime}\Big|_\text{\Reff{Buras:2000if}}
=
\left(
\setlength{\arraycolsep}{4.5pt}
\begin{array}{ccccccccc}
0 & 0 & -\frac43 & 4 & -\frac43 & 4 & 0 & \cdots & 0 \\
0 & 0 & -\frac43 & 4 & -\frac43 & 4 & 0 & \cdots & 0 \\
\multicolumn{9}{c}{0_{6\times 18}} \\
0 & 0 & -2 & 6 & -2 & 6 & 0 & \cdots & 0 \\
0 & 0 & -2 & 6 & -2 & 6 & 0 & \cdots & 0 \\
\multicolumn{9}{c}{0_{8\times 18}} 
\end{array}
\right)\ .
}
We therefore conclude that there is a problem with the matrix $\gamma^{(1)}_\text{BMU}$ as given in~\Reff{Buras:2000if}, most likely related to penguin contributions, in the BSM sector.

%%%%%%%%%%%%%%%%%%%%%%%%%%%%%%%%%%%%%%%%%%%%%%%
\section{The Diagnosis: a naive treatment of $Q_{11}$}
\label{sec:diagnosis}

\subsection{The original approach in BJLW and BMU}

Anomalous dimensions in dimensional regularization can be calculated setting the quark masses to zero, given that they depend exclusively on the UV structure of the theory.
This means that, up to quark-mass effects, one has for example
\eq{
\label{symmetrydiagrams}
\text{Penguin\ diagram}(Q_3) =
f\cdot \text{Penguin\ diagram}(\widetilde Q_1) 
+2\cdot \text{Penguin\ diagram}(Q_2)\ ,
}
where the first term in the RHS proportional to the number of active flavors $f$ accounts for closed penguins, and the second term accounts for the two open penguins with $s$ and $d$ quarks in the loop.
This sort of relations allows one to take a calculation involving insertions of a certain reduced set of operators and extend them to infer the calculations involving a full operator basis.

This methodology was used by BJLW in~\Reff{Buras:1992tc} to compute the  $\cO(\alpha_s^2)$ contributions to the ADM for the ten SM operators, and later in~\Reff{Buras:2000if} for the full set of forty operators in the general BSM case (the BMU basis, see~\App{app:OPBasis}).
In both cases the corresponding ADMs were built out of a small set of tables of pole coefficients computed in~Refs.~\cite{Buras:1989xd, Buras:1992tc, Buras:2000if} for a single quark flavor. Of all the contributions considered in~Refs.~\cite{Buras:1992tc,Buras:2000if}, we shall focus exclusively on the ones coming from penguin diagrams with insertions of VLL operators, as discussed above. 

The building blocks for the NLO VLL-penguin ADM are the tables of two-loop pole-coefficients computed in~\Reff{Buras:1992tc} for $Q_1$ and $Q_2$, which involve only \textit{open} penguin diagrams.
\Reff{Buras:1992tc} proceeds then by performing a change of basis into a basis where the first two operators ($\widetilde{Q}_1$ and $\widetilde{Q}_2$) are modified to be \textit{penguin-closed} (i.e. with the structure $\bar{s}b\bar{u}u$ instead of $\bar{s}u\bar{u}b$). Thus, four separate contributions to the anomalous dimensions are obtained: $[\hat{\gamma}^{(1)}(Q_1)]_p, [\hat{\gamma}^{(1)}(Q_2)]_p, [\hat{\gamma}^{(1)}(\widetilde{Q}_1)]_p$ and $[\hat{\gamma}^{(1)}(\widetilde{Q}_2)]_p$. 
Once these basic ingredients are known, \Reff{Buras:1992tc} proceeds by taking advantage of flavor-independence of the various Feynman diagrams (e.g.~\Eq{symmetrydiagrams}), and reconstructing the penguin contributions of all VLL penguin operators ($Q_3, Q_4, Q_9, Q_{10}$) simply by combining the only four independent pieces,
\eqa{
\label{ADM_Q3}
\Big[\hat{\gamma}^{(1)}(Q_{3})\Big]_p &\quad=\quad& 
f \Big[\hat{\gamma}^{(1)}(\widetilde{Q}_{1})\Big]_p + 2 \Big[\hat{\gamma}^{(1)}(Q_{2})\Big]_p 
\ , 
\\
\label{ADM_Q4}
\Big[\hat{\gamma}^{(1)}(Q_{4})\Big]_p &\quad=\quad&
f \Big[\hat{\gamma}^{(1)}(\widetilde{Q}_{2})\Big]_p + 2 \Big[\hat{\gamma}^{(1)}(Q_{1})\Big]_p 
\ , 
\\
\label{ADM_Q9}
\Big[\hat{\gamma}^{(1)}(Q_{9})\Big]_p &\quad=\quad&
(u Q_u + d Q_d) \Big[\hat{\gamma}^{(1)}(\widetilde{Q}_{1})\Big]_p + 2 \,Q_d \Big[\hat{\gamma}^{(1)}(Q_{2})\Big]_p 
\ , 
\\
\label{ADM_Q10}
\Big[\hat{\gamma}^{(1)}(Q_{10})\Big]_p &\quad=\quad&
(u Q_u + d Q_d) \Big[\hat{\gamma}^{(1)}(\widetilde{Q}_{2})\Big]_p + 2 \,Q_d \Big[\hat{\gamma}^{(1)}(Q_{1})\Big]_p 
\ .
}
These relations involve anomalous dimensions, and not just Feynman diagrams as in~\Eq{symmetrydiagrams}, and thus the extra terms in the RHS come from an additional assumption for $s$ and $b$ quarks: that one can get these special cases (which contribute simultaneously via open and closed penguin diagrams) through the separate combination of open and closed penguins,
\eqa{
\label{VSLL}
\Big[\hat{\gamma}^{(1)}(\mathcal{O}^{VS,LL}_{sdss})\Big]_p &=
\Big[\hat{\gamma}^{(1)}(\mathcal{O}^{VS,LL}_{sddd})\Big]_p =&
\Big[\hat{\gamma}^{(1)}(\widetilde{Q}_{1})\Big]_p + \Big[\hat{\gamma}^{(1)}(Q_2)\Big]_p 
\ , 
\\
\label{VXLL}
\Big[\hat{\gamma}^{(1)}(\mathcal{O}^{VX,LL}_{sdss})\Big]_p &=
\Big[\hat{\gamma}^{(1)}(\mathcal{O}^{VX,LL}_{sddd})\Big]_p =&
\Big[\hat{\gamma}^{(1)}(\widetilde{Q}_{2})\Big]_p + \Big[\hat{\gamma}^{(1)}(Q_1)\Big]_p 
\ ,
}
where the generic operators $\mathcal{O}^{A,B}_{ijkl}$ are defined at the end of~\App{app:OPBasis}. 

In their posterior work, BMU derive the anomalous dimensions for the BSM operators $Q_{11,12,13}$ in a similar way.
While \Reff{Buras:2000if} is not completely explicit on the exact procedure followed and on the evanescent operator basis used for this sector, it does literally state that: 
(A) the current-current contributions can be directly taken from the ADMs for $\Delta F=2$ and $\Delta F=1$ operators of the type $(\bar s u)(\bar c d)$, and
(B) the penguin contributions can be ``easily'' extracted from Sections~3.2 and~5.3 of~\Reff{Buras:1992tc}. 
From statement (A) we infer that the evanescent basis is equivalent to the one used here (see~\App{app:EVBasis}), and we confirm their results for current-current contributions.
From statement (B) we infer that the penguin contributions are obtained from the following relations,\footnote{
We thank Mikolaj Misiak for confirming to us that this was indeed the approach followed in~\Reff{Buras:2000if}.
}
\eqa{
\label{gammaQ11}
\Big[\hat{\gamma}^{(1)}(Q_{11})\Big]_p &\quad=\quad& 
{\Big[\hat{\gamma}^{(1)}(Q_{3})\Big]_p}\;\Big|_{f=2} 
\qquad\text{(allegedly)}
\ , \\
\label{gammaQ12}
\Big[\hat{\gamma}^{(1)}(Q_{12})\Big]_p &\quad=\quad& {\Big[\hat{\gamma}^{(1)}(Q_{6})\Big]_p}\;\Big|_{f=2} 
\qquad\text{(allegedly)}
\ , \\
\label{gammaQ13}
\Big[\hat{\gamma}^{(1)}(Q_{13})\Big]_p &\quad=\quad& {\Big[\hat{\gamma}^{(1)}(Q_{5})\Big]_p}\;\Big|_{f=2}  
\qquad\text{(allegedly)}
\ , 
}
where $f=2$ indicates a calculation with only two active quark flavors ($d$ and $s$). 
These relations are all again presumably inspired by the (correct) statement in~\Eq{symmetrydiagrams}, and result from the application of~Eqs.~(\ref{VSLL}) and~(\ref{VXLL}) to $Q_{11-13}$.
We can confirm that using~Eqs.~(\ref{gammaQ11})-(\ref{gammaQ13}) we reproduce the LO and NLO ADMs given by BMU.

The relation for $Q_{11}$ in~\Eq{gammaQ11} can be combined with~\Eq{ADM_Q3} and rewritten as
\eqa{
\label{BMU_VSLL}
\Big[\hat{\gamma}^{(1)}(Q_{11})\Big]_p &\quad=\quad& 
2 \Big[\hat{\gamma}^{(1)}(\widetilde{Q}_{1})\Big]_p + 2 \Big[\hat{\gamma}^{(1)}(Q_2)\Big]_p 
\qquad\text{(allegedly)}
\ .
}
The BMU ADMs also satisfy this relation.
Our claim here is that, while~Eqs.~(\ref{VSLL}) and~(\ref{VXLL}) are true when used within~Eqs.(\ref{ADM_Q3})-(\ref{ADM_Q10}) in the set of operators $\{Q_3, Q_4, Q_9, Q_{10}\}$ of the SM sector, the approach fails in~\Eq{BMU_VSLL} as used in~\Reff{Buras:2000if}, for $Q_{11}$ alone. The key point to understand our claim lies in the intermediate one-loop contributions participating in the ADM, coming from the insertion of one-loop counterterms in the divergent subdiagrams of two-loop penguins. These terms end up providing a contribution that depends not only on the operator inserted in the two-loop diagram, but also on a closed set of operators around it. In particular, we will see how the contribution from the one-loop counterterms to $\{Q_1, Q_2\}$ and $\{\widetilde{Q}_1, \widetilde{Q}_2\}$ cannot be used directly to recover the one they provide for $Q_{11}$, regardless of flavor symmetry.

%%%%%%%%%%%%%%%%%%%%%%%%%%%%%%%%%%%%%%%%%%%%%%%
\subsection{Deconstruction of~\Eq{BMU_VSLL}}

We start from the expression for the two-loop ADM in~\Eq{gamma1a},
focusing only on the penguin contributions,
\eqa{
\label{2loopADM_a}
\Big[\hat{\gamma}^{(1)}(Q_i)_j\Big]_p &=&
-4 \Big[\hat{a}^{(2,1)}_{Q_iQ_j}
\Big]_p + 4 \Big[ \hat{a}^{(1,1)}_{Q_iQ_k}\hat{a}^{(1,0)}_{Q_kQ_j} \Big]_p + 2 \Big[\hat{a}^{(1,1)}_{Q_iE_k}\hat{a}^{(1,0)}_{E_kQ_j}\Big]_p 
\ .
}
The penguin brackets $[...]_p$ indicate that only the contributions that involve at least one penguin diagram are considered. \Eq{2loopADM_a} allows for a closer inspection on the source of all the different contributions and their role in~\Eq{BMU_VSLL}:

\subsubsection*{\bf First term in the RHS of~\Eq{2loopADM_a}}

The first term in the RHS of~\Eq{2loopADM_a} comes from $1/\epsilon$ poles in the bare one- and two-loop penguin diagrams. This contribution projects always only onto $Q_{3-6}$~\cite{Buras:2000if} and depends only on the definition of $Q_i$. It is also clearly independent of the flavor of the quark in the loop. Therefore, it allows for~\Eq{BMU_VSLL} to be applied without further dependence on the context.

It is then clear that if there is to be some dependence on intermediate operators that spoils the validity of~\Eq{BMU_VSLL}, it must come from a physical $Q_k$ as in the second term in~\Eq{2loopADM_a}, or from an evanescent $E_k$ in the third term.

\subsubsection*{\bf Second term in the RHS of~\Eq{2loopADM_a}}

We can separate this term into three contributions, depending on the type of diagrams involved,
\eqa{
\label{a_prod}
\Big[ \hat{a}^{(1,1)}_{Q_iQ_k}\hat{a}^{(1,0)}_{Q_kQ_j} \Big]_p &=& \Big[\hat{a}^{(1,1)}_{Q_iQ_k}\Big]_{cc} \Big[\hat{a}^{(1,0)}_{Q_kQ_j}\Big]_p + \Big[\hat{a}^{(1,1)}_{Q_iQ_k}\Big]_p \Big[\hat{a}^{(1,0)}_{Q_kQ_j}\Big]_{cc} + \Big[\hat{a}^{(1,1)}_{Q_iQ_k}\Big]_p \Big[\hat{a}^{(1,0)}_{Q_kQ_j}\Big]_p 
\ .
}
Among the various terms in~\Eq{a_prod}, those containing $[\hat{a}^{(1,1)}_{Q_iQ_k}]_p$ involve (at most) only $Q_k = Q_{3-6}$ as intermediate operators, for any $Q_i$ inserted. Therefore, this term provides universal contributions too, and again allows for a separate use of the naive relation in~\Eq{BMU_VSLL}. 

This is not the case, however, for the term containing $[\hat{a}^{(1,1)}_{Q_iQ_k}]_{cc}$, in which $Q_k$ runs only through the set of operators connected to $Q_i$ by one-loop current-current diagrams. This set is a pair of color-singlet and color-crossed operators for $Q_i = Q_1,Q_2$ and their \textit{tilde} versions. Meanwhile, for $Q_i = Q_{11}$ one has $Q_k = Q_{11}$, featuring only a color-singlet. The contributions in both sides of~\Eq{BMU_VSLL} read then, up to an overall factor of 8,
\begin{align*}
    \text{LHS} \;: \quad \Big[\hat{a}^{(1,1)}_{Q_2Q_1}\Big]_{cc}\Big[\hat{a}^{(1,0)}_{Q_2Q_j}\Big]_p + \Big[\hat{a}^{(1,1)}_{Q_2Q_2}\Big]_{cc}\Big[\hat{a}^{(1,0)}_{Q_2Q_j}\Big]_p + 
    \Big[\hat{a}^{(1,1)}_{\widetilde{Q}_1\widetilde{Q}_1}\Big]_{cc}\Big[\hat{a}^{(1,0)}_{\widetilde{Q}_1Q_j}\Big]_p + 
    \Big[\hat{a}^{(1,1)}_{\widetilde{Q}_1\widetilde{Q}_2}\Big]_{cc}\Big[\hat{a}^{(1,0)}_{\widetilde{Q}_1Q_j}\Big]_p \;, \\
    \text{RHS} \;: \quad \Big[\hat{a}^{(1,1)}_{Q_{2}Q_{1}}\Big]_{cc}\Big[\hat{a}^{(1,0)}_{Q_1Q_j}\Big]_p + \Big[\hat{a}^{(1,1)}_{Q_2Q_2}\Big]_{cc}\Big[\hat{a}^{(1,0)}_{Q_2Q_j}\Big]_p + 
    \Big[\hat{a}^{(1,1)}_{\widetilde{Q}_1\widetilde{Q}_1}\Big]_{cc}\Big[\hat{a}^{(1,0)}_{\widetilde{Q}_1Q_j}\Big]_p + 
    \Big[\hat{a}^{(1,1)}_{\widetilde{Q}_1\widetilde{Q}_2}\Big]_{cc}\Big[\hat{a}^{(1,0)}_{\widetilde{Q}_2Q_j}\Big]_p \;, 
\end{align*}
where we have used the fact that the $1/\epsilon$ poles in one-loop diagrams are scheme-independent to write all of the corresponding matrices in terms of the two $u$-type operators. We have also taken into account that 
\eqa{
\left[\hat{a}^{(1,0)}_{Q_{11}Q_j}\right]_p &\quad=\quad& 
2\left[\hat{a}^{(1,0)}_{\widetilde{Q}_1Q_j}\right]_p + 2\left[\hat{a}^{(1,0)}_{Q_{2}Q_j}\right]_p 
\ ,
}
which is only the one-loop statement that $Q_{11}$ contributes both through closed and open penguin diagrams. 
It is readily apparent that the LHS and RHS of~\Eq{BMU_VSLL} differ in the first and last terms. Numerically, written in terms of $Q_j = (Q_3, Q_4, Q_5, Q_6)$, the difference (factor of 8 included) amounts to 
\eqa{
\label{RHS-LHS_Phys}
\text{LHS} - \text{RHS} \;\Big|_{\text{second\ term}} &\quad=\quad& 
8\,\Big(
\begin{array}{ccccccc}
\frac{1}{N_c} && -1 && \frac{1}{N_c} && -1 \end{array}
\Big) 
\ ,
}
computed in the renormalization scheme defined below~\Eq{ubsym}. This non-zero result does not pose any problem {\it per se}, as it could cancel against the third term of~\Eq{2loopADM_a}.

\subsubsection*{\bf Third term in the RHS of~\Eq{2loopADM_a}}

There is a similar situation for the evanescent contribution in~\Eq{2loopADM_a}, further simplified by the fact that one-loop penguin insertions of physical operators produce no evanescent structures. Therefore, only the current-current $1/{\epsilon}$ poles will contribute. Given that the set of evanescent operators are defined independently of the physical basis, as long as they respect quark-flavor symmetry (analogous evanescents for each flavor) the contribution from the third term in~\Eq{2loopADM_a} to each $[\hat{\gamma}^{(1)}(Q_i)]_p$ will be flavor-universal, and thus have $\text{LHS} = \text{RHS}$ in~\Eq{BMU_VSLL}.
This is indeed the case of the evanescent basis used by BMU, as argued below~\Eq{ubsym}.

Nonetheless, for the special case of $Q_{11}$ there is an additional evanescent structure with no analog associated to $Q_{1,2}$ or $\widetilde Q_{1,2}$, needed in the one-loop current-current diagrams with an insertion of $Q_{11}$,
\eq{
\label{E_11}
E_{11} \equiv Q'_{11} - Q_{11} = E^{\text{VLL}(d)}_1 + E^{\text{VLL}(s)}_1 \;.
}
The leftmost equality in~\Eq{E_11} is written as in \Reff{Buras:2000if} (cf.~\App{app:OPBasis} for the definition of these operators), while the rightmost expression is written in terms of the evanescent operators listed in \App{app:EVBasis}. Due to the emergence of this evanescent structure, the LHS of~\Eq{BMU_VSLL} gets an additional contribution that is not present in the RHS, given by
\eqa{
\label{RHS-LHS_Ev}
\text{LHS} - \text{RHS} \;\Big|_{\text{third\ term}} &\;=\;
2\Big[\hat{a}^{(1,1)}_{Q_{11}E_{11}}\Big]_{cc}\Big[\hat{a}^{(1,0)}_{E_{11}Q_j}\Big]_p \;=\;& 
4\,\Big(
\begin{array}{ccccccc}
-\frac{1}{N_c} && 1 && -\frac{1}{N_c} && 1 
\end{array}
\Big) 
\ .
}

\subsection{Correction to~\Eq{BMU_VSLL}}

Putting together the two contributions in~Eqs.~(\ref{RHS-LHS_Phys}) and~(\ref{RHS-LHS_Ev}), we can write
\eqa{
\label{BMU_VSLL_correct}
\Big[\hat{\gamma}^{(1)}(Q_{11})\Big]_p &\quad=\quad& 
2 \Big[\hat{\gamma}^{(1)}(\widetilde{Q}_{1})\Big]_p + 2 \Big[\hat{\gamma}^{(1)}(Q_2)\Big]_p 
+ \Delta_{11}
\ ,
}
with
\eqa{
\label{RHS-LHS}
\Delta_{11} & = &
\Big(
\begin{array}{ccccccccccccccccc}
0 && 0 && \frac4{N_c} && -4 && \frac4{N_c} && -4
&& 0 && \cdots && 0
\end{array}
\Big) 
\ .
}
This correction is the reason behind the inconsistency found in the NLO ADM given in~\Reff{Buras:2000if}, as discussed in~\Sec{sec:problem}, and it is contained entirely in the anomalous dimension of the BSM operator $Q_{11}$.

%%%%%%%%%%%%%%%%%%%%%%%%%%%%%%%%%%%%%%%%%%%%%%%
\section{The Solution}
\label{sec:solution}

Applying this correction to the ADM of~\Reff{Buras:2000if} we get, for the 11th row of $\hat \gamma^{(1)}$,
\eqa{
\gamma^{(1)}_{11\,j} & = &
\gamma^{(1)}_{11\,j}\Big|_\text{\Reff{Buras:2000if}} + \Delta_{11}
\nonumber\\[2mm]
&&
\label{correctQ11}
= \Big(
\begin{array}{ccccccccccccccccccccccccccc}
0 && 0 && \red{\frac{3862}{243}} && \red{\frac{2330}{81}} && \red{-\frac{5894}{243}} && \red{\frac{1430}{81}} && 0 && 0 && 0 && 0 
&& \frac{4f}{9}-7 && 0 && \cdots && 0
\end{array}
\Big) 
\ ,
}
where $f$ indicates the number of quark flavors, and we have set $N_c=3$ for simplicity. The general expression in terms of $N_c$ is given below in~\Sec{sec:proposal}. We have also indicated in \red{red} the four terms that are different from~\Reff{Buras:2000if}.

With our corrected version of $\hat\gamma^{(1)}$ at hand we can now verify that~\Eq{ubsym} is, indeed, satisfied. That is,
\eq{
\hat{\gamma}^{(1)}_\text{\Reff{Buras:2000if}} 
- \hat{R} \hat{\gamma}^{(1)}_\text{\Reff{Buras:2000if}} \hat{R}^{-1}
+ 2\beta_0 \Delta\hat{r}
+ \Big[\Delta\hat{r}, \hat{\gamma}^{(0)}_{\text{BMU}}\Big]
=
\hat{R} \Delta_{11} \hat{R}^{-1}
- \Delta_{11}
\ ,
}
as can be checked explicitly by noting that the right-hand-side agrees exactly with the matrix in~\Eq{BMU-BMU'}.
(Here we have made a slight abuse of notation by denoting by $\Delta_{11}$ the matrix with $\Delta_{11}$ as the 11th row and all other entries vanishing.)
Thus we are confident that the diagnosis in the previous section is correct, and that no other issues, aside from the one related to $Q_{11}$, affect the results of~\Reff{Buras:2000if}.

Our results can also be compared to the results for the anomalous dimensions of the operator $P_b$ in~\Reff{Huber:2005ig} (adjusting for the case of our $\bar s\to \bar d$ transition),
\eq{
P_b = \frac1{12} (\bar s^\alpha \gamma^\mu\gamma^\nu\gamma^\rho P_L d^\alpha)(\bar d^\beta \gamma_\mu\gamma_\nu\gamma_\rho d^\beta)
-\frac13 (\bar s^\alpha \gamma^\mu P_L d^\alpha)(\bar d^\beta \gamma_\mu d^\beta)\ ,
}
and in particular to the two-loop mixing of $P_b$ onto the QCD penguin operators $P_3 - P_6$
\eq{
\label{gammaHuber}
\gamma^{(20)}_{BP} = \bigg(
-\frac{1576}{81}\quad
\frac{446}{27}\quad
\frac{172}{81}\quad
\frac{40}{27}\bigg)\ .
}
In the BMU basis, the operator $P_b$ is given by
\eq{
P_b = 
\frac1{12}\bigg[ 
(6 - 2\epsilon) (Q_{11} + Q_{14}) 
+ 2\epsilon\,(Q_{13}+Q_{16})
+ E_2^{\text{VLL}(d)}
+ E_2^{\text{VLR}(d)}
\bigg]\ .
}
We perform a change of basis from the BMU basis to the basis of~\Reff{Huber:2005ig} (taking into account that a different basis for evanescent operators is used in that paper), and we confirm the anomalous dimensions in~\Eq{gammaHuber}, only when using the new results in~\Eq{correctQ11}.

As a final note, we note that our results in~\Eq{correctQ11} have been confirmed a posteriori in an erratum to~\Reff{Buras:2000if}.

%%%%%%%%%%%%%%%%%%%%%%%%%%%%%%%%%%%%%%%%%%%%%%%
\section{A Proposal: Crossed/Singlet symmetrization}
\label{sec:proposal}

The rationale behind this discrepancy is the fact that the four VLL penguin contributions to the ADM computed in~\Reff{Buras:1992tc} (for $Q_1$, $Q_2$ and their tildes) are valid only for cases with an analogous set of operators connected by one-loop current-current diagrams, which should involve a pair of color-singlet and color-crossed operators. If we want to extrapolate these results to $d$-type and $s$-type operators, we must then use a properly crafted operator that is connected to an equivalent set. Such property can be found, for instance, in a modified version of $Q_{11}$ that symmetrizes over color structures,
$Q_{11}^+ = \frac{1}{2} \, Q_{11} + \frac{1}{2}\, \widetilde{Q}_{11}$.
The connected set for this operator is again only itself, but it now includes the proper pair of singlet/crossed structures, with which the counterterm contributions become
\eqa{
\left[\hat{a}^{(1,1)}_{Q_{11}^+Q_{11}^+}\right]_{cc} & =  \left[\hat{a}^{(1,1)}_{Q_2Q_1}\right]_{cc} + \left[\hat{a}^{(1,1)}_{Q_2Q_2}\right]_{cc} + \left[\hat{a}^{(1,1)}_{Q_1Q_1}\right]_{cc} + \left[\hat{a}^{(1,1)}_{Q_1Q_2}\right]_{cc}  = & \left[\hat{a}^{(1,1)}_{Q_{11}Q_{11}}\right]_{cc} \;, \\ 
\left[\hat{a}^{(1,0)}_{Q_{11}^+Q_j}\right]_p & =  \left[\hat{a}^{(1,0)}_{\widetilde{Q}_1Q_j}\right]_p + \left[\hat{a}^{(1,0)}_{Q_{2}Q_j}\right]_p + \left[\hat{a}^{(1,0)}_{\widetilde{Q}_2Q_j}\right]_p + \left[\hat{a}^{(1,0)}_{Q_{1}Q_j}\right]_p  \neq & \left[\hat{a}^{(1,0)}_{Q_{11}Q_j}\right]_p
\ . 
}
The product of these two expressions now aligns perfectly with the decomposition in terms of operators $Q_1, Q_2, \widetilde{Q}_1$ and $\widetilde{Q}_2$,
\eq{
\begin{aligned}
\left[\hat{a}^{(1,1)}_{Q_{11}^+Q_{11}^+}\right]_{cc} \left[\hat{a}^{(1,0)}_{Q_{11}^+Q_j}\right]_p \; = & \quad\left[\hat{a}^{(1,1)}_{Q_{2}Q_{k}}\right]_{cc} \left[\hat{a}^{(1,0)}_{Q_{k}Q_j}\right]_p + \left[\hat{a}^{(1,1)}_{\widetilde{Q}_1Q_{k}}\right]_{cc} \left[\hat{a}^{(1,0)}_{Q_{k}Q_j}\right]_p \\ 
 & + \left[\hat{a}^{(1,1)}_{Q_{1}Q_{k}}\right]_{cc} \left[\hat{a}^{(1,0)}_{Q_{k}Q_j}\right]_p + \left[\hat{a}^{(1,1)}_{\widetilde{Q}_2Q_{k}}\right]_{cc} \left[\hat{a}^{(1,0)}_{Q_{k}Q_j}\right]_p 
\ .
\end{aligned}
}
In the evanescent plane of~\Eq{RHS-LHS_Ev}, $Q_{11}^+$ has two identical and opposite-sign contributions to $[\hat{a}^{(1,1)}_{Q_{11}E_{11}}]_p$, given that insertions of color-crossed operators project onto $-E_{11}$; and thus this contribution to the discrepancy between the actual contribution and its construction from single-flavor results vanishes too for $Q_{11}^+$ (that is, $\Delta^+_{11} = 0$). 

With both the physical and evanescent contributions to the ADM agreeing for $Q_{11}^+$ on the naive comparison with $u$-type operators, we can now safely apply the respective naive reconstruction of the penguin-borne anomalous dimension,
\eqa{
\label{VSLL+VXLL}
\Big[\hat{\gamma}^{(1)}(Q_{11}^+)\Big]_p &\quad=\quad&
\Big[\hat{\gamma}^{(1)}(\widetilde{Q}_{1})\Big]_p + \Big[\hat{\gamma}^{(1)}(Q_2)\Big]_p + \Big[\hat{\gamma}^{(1)}(\widetilde{Q}_{2})\Big]_p + \Big[\hat{\gamma}^{(1)}(Q_1)\Big]_p 
\ .
}
One can then perform a NLO change of basis from this quasi-BMU basis containing $Q_{11}^+$ to the original BMU basis, to obtain the correct $[\hat{\gamma}^{(1)}(Q_{11})]_p$. This change of basis affects only $Q_{11}$, and leaves the rest of the ADM (and in particular the SM sector) unaltered. The resulting contributions from either operator to the ADM read
\eqa{
\label{ADM_Q11+}
\Big[\hat{\gamma}^{(1)}(Q_{11}^+)\Big]_p \quad &= \quad 
\left(\begin{array}{c}
    \frac{160}{27 N_c^2}+6 N_c-\frac{10}{3 N_c}-\frac{52}{27} \\ \frac{286 N_c}{27}-\frac{394}{27 N_c}-\frac{8}{3} \\ -\frac{92}{27 N_c^2}-6 N_c+\frac{26}{3 N_c}-\frac{178}{27} \\ \frac{160 N_c}{27}+\frac{110}{27 N_c}-\frac{8}{3} 
\end{array}\right)^T 
\ , 
\\
\label{ADM_Q11}
\left[\hat{\gamma}^{(1)}(Q_{11})\right]_p \quad &= \quad 
\left(\begin{array}{c}
    \frac{172}{27 N_c^2}+6 N_c-\frac{4}{3 N_c}-\frac{64}{27} \\ \frac{352 N_c}{27}-\frac{460}{27 N_c}-\frac{14}{3} \\ -\frac{188}{27 N_c^2}-6 N_c+\frac{32}{3 N_c}-\frac{244}{27} \\ \frac{172 N_c}{27}+\frac{260}{27 N_c}-\frac{14}{3}
\end{array}\right)^T
\ , 
} 
with these vectors being written in terms of the four QCD penguins $(Q_3,\, Q_4,\, Q_5,\, Q_6)$.
\Eq{ADM_Q11} is the corrected version of the penguin contribution to the ADM due to $Q_{11}$, and agrees with the result given in~\Sec{sec:solution} for $N_c=3$.

Going back to our original claim below~\Eq{BMU_VSLL}, we can see that, as opposed to~\Eq{BMU_VSLL}, Eqs.~(\ref{VSLL}) and (\ref{VXLL}) are correct because the penguin operators $Q_{3}, Q_{4}, Q_{9}, Q_{10}$ are built respecting the required structure of color-singlet/crossed pairs. 
Consequently, one is allowed to directly export the single-flavor penguin anomalous dimensions as in~Eqs.~(\ref{ADM_Q3}), (\ref{ADM_Q4}), (\ref{ADM_Q9}) and~(\ref{ADM_Q10}), leading to the results given in~\Reff{Buras:1992tc}, which are in full agreement with multiple independent calculations of the anomalous dimensions at $O(\alpha_s^2)$ performed for the SM sector \cite{Gorbahn:2004my, Ciuchini:1993vr, Ciuchini:1993fk, Bobeth:2003at, Huber:2005ig}, after the proper change of basis.

%%%%%%%%%%%%%%%%%%%%%%%%%%%%%%%%%%%%%%%%%%%%%%%
\section{Numerical impact of the correction}
\label{sec:numerics}

We now study the phenomenological impact of the correction put forward in this work.
We do this by comparing the Renormalization Group Evolution resulting from BMU on the one hand, and from our results on the other. 
We compute the running between two representative scales, from $\mu_0 \sim M_Z$ (i.e. the scale of a matching to the SMEFT) to $\mu \sim m_b$ (the characteristic scale of $B$-physics).

Limiting ourselves to contributions of dimension 6, i.e. of order $1/{\Lambda^2}$, the mixing relevant to penguin operators involves only single insertions of the first thirteen operators in the BMU basis (c.f.~\App{app:OPBasis}). In this situation the equation for the running can be written in terms of the unitary evolution matrix,
\eq{
\label{RGE_Matrix}
    C_i(\mu) = \hat{U}_{ij}(\mu,\mu_0)\,C_j(\mu_0) \;.
}
This matrix can then be computed as the solution to the RGE in~\Eq{RGE}, with the appropriate boundary conditions. The general solution reads:
\eq{
\label{Umatrix}
    \hat{U}(\mu,\mu_0) = \exp{\left(\int_{\mu_0}^{\mu}\hat{\gamma}(\mu')\, d\log{\mu'}\right)} = \exp{\left(\int_{\alpha_s(\mu_0)}^{\alpha_s(\mu)}\frac{\hat{\gamma}(\alpha_s)}{2\beta(\alpha_s)}\,\frac{d\alpha_s}{\alpha_s}\right)} \;,
}
where the anomalous dimensions, $\hat{\gamma}$, and the QCD beta function, $\beta$, can be expanded perturbatively in $\alpha_s$.
Solving the RGE numerically to NLO both in the ADM and the QCD beta function, one obtains the corresponding $13\times13$ matrix, 
\eq{
\label{U(Mz,mb)}
\begin{aligned}
    &\hat{U}(m_b,M_Z) = \\ &\left(
\begin{array}{rrrrrrrrrrrrr}
 1.11 & -0.24 & 0 & 0 & 0 & 0 & 0 & 0 & 0 & 0 & 0 & 0 & 0 \\
 -0.24 & 1.11 & 0 & 0 & 0 & 0 & 0 & 0 & 0 & 0 & 0 & 0 & 0 \\
 -0.01 & 0.01 & 1.11 & -0.19 & 0.02 & 0.08 & 0 & 0.01 & -0.01 & 0.01 & 0.01 & 0.03 & 0.01 \\
 0.01 & -0.03 & -0.28 & 0.97 & -0.01 & -0.17 & 0 & -0.02 & 0.03 & -0.02 & -0.05 & -0.07 & -0.01 \\
 0 & 0.01 & 0.03 & 0.04 & 0.92 & 0.09 & 0 & 0 & -0.01 & 0 & 0.02 & 0.01 & -0.01 \\
 0.01 & -0.04 & -0.05 & -0.16 & 0.32 & 1.71 & 0 & -0.02 & 0.04 & -0.02 & -0.06 & -0.10 & -0.01 \\
 0 & 0 & 0 & 0 & 0 & 0 & 0.93 & 0.06 & 0 & 0 & 0 & 0 & 0 \\
 0 & 0 & 0 & 0 & 0 & 0 & 0.34 & 1.95 & 0 & 0 & 0 & 0 & 0 \\
 0 & 0 & 0 & 0 & 0 & 0 & 0 & 0 & 1.11 & -0.24 & 0 & 0 & 0 \\
 0 & 0 & 0 & 0 & 0 & 0 & 0 & 0 & -0.24 & 1.11 & 0 & 0 & 0 \\
 0 & 0 & 0 & 0 & 0 & 0 & 0 & 0 & 0 & 0 & 0.87 & 0 & 0 \\
 0 & 0 & 0 & 0 & 0 & 0 & 0 & 0 & 0 & 0 & 0 & 1.95 & 0.34 \\
 0 & 0 & 0 & 0 & 0 & 0 & 0 & 0 & 0 & 0 & 0 & 0.06 & 0.93 \\
\end{array}
\right) .
\end{aligned}
}
The correction to the NLO ADM affects only the entries mixing $Q_{11}$ into $Q_3-Q_6$ (rows third to sixth in the eleventh column). Focusing on these entries (to a precision of four significant figures, consistent with an $\hat \alpha_s(m_b)^2$ correction) and comparing them to the calculation with the original ADM of~\Reff{Buras:2000if}, one finds
\eqa{
\label{U(Mz,mb)_focus}
\begin{aligned}
    &\left[\hat{U}^{(\text{this paper})}(m_b,M_Z)\right]_{i\,11} = \left(
    \begin{array}{c}
     0.0127 \\
     -0.0534 \\
     0.0206 \\
     -0.0619 \\
    \end{array}
    \right) , \quad
    &\left[\hat{U}^{\text{(\Reff{Buras:2000if})}}(m_b,M_Z)\right]_{i\,11} = \left(
\begin{array}{c}
 0.0134 \\
 -0.0550 \\
 0.0211 \\
 -0.0639 \\
\end{array}
\right)\, .
\quad
\end{aligned}
}
The difference in these entries is of the order of $5\%$. Although small in absolute terms, the impact of such corrections could become sizeable in phenomenological studies where the BSM matching condition $C_{11}(M_Z)$ is significantly larger than the SM contribution to QCD penguins, $C_{3-6}(M_Z)$. In such cases, the running described by~\Eq{U(Mz,mb)} could lead to similar contributions by both SM and BSM to the coefficients $C_{3-6}(m_b)$ at the low scale, and the corrections in~\Eq{U(Mz,mb)_focus} would then make a measurable difference to suitable observables. It remains to be clarified to which extent current data allows for large values of $C_{11}(M_Z)$.

%%%%%%%%%%%%%%%%%%%%%%%%%%%%%%%%%%%%%%%%%%%%%%%
\section{Summary}
\label{sec:conclusion}

In this paper we have revisited the two-loop anomalous dimensions for $\Delta F=1$ four-quark operators in the general BSM case.
These anomalous dimensions were presented in complete form for the first time in the highly relevant paper by Buras, Misiak and Urban (BMU) in the year 2000~\cite{Buras:2000if}.
However, the BMU result for the NLO anomalous dimension matrix $\hat \gamma^{(1)}$ does not satisfy a simple requirement related to renaming of quark fields.

The root of the problem is related to the particular structure of the operator $Q_{11}$, 
an issue that, once addressed, can be used to derive the correct version of the anomalous dimensions, which can be found in~\App{app:BMU_ADM}. Our corrected version satisfies the renaming requirement, and thus confirms our diagnosis of the problem.
Having understood the issue, the approach followed by BMU can be modified in a way that leads directly to the correct result.
Our findings have been confirmed by the authors of~\Reff{Buras:2000if}, and the BMU results have been extended to derive the full set of two-loop anomalous dimensions for all four-fermion dimension-six operators in the LEFT~\cite{Aebischer:2025hsx}.

In order to assess the numerical importance of this correction, we have performed a very simple numerical analysis that points to an effect of around $\sim 5\%$.
Our results are also very relevant in the present time in which automation is prompting the development of public codes which implement computations in EFTs in full generality~\cite{Aebischer:2023nnv,Celis:2017hod,Fuentes-Martin:2020zaz,Aebischer:2018bkb,EOSAuthors:2021xpv}.

Many of the points put forward in this work can be applied to general $n$-loop anomalous dimensions. On the one hand, as long as the evanescent basis is properly defined, quark-flavor symmetry tests are completely general consistency checks. On the other hand, analyses like the one carried out in \Sec{sec:diagnosis} are always necessary when trying to extend calculations performed in small operator subsets to other sectors of the basis.
One must ensure that both sectors have analogous physical and evanescent ``surroundings'', as the direct extension fails otherwise.
It is possible that the issues discussed in this paper can be framed within recent attempts to simplify the handling of evanescent structures in loop calculations~\cite{Aebischer:2022tvz,Aebischer:2022aze,Aebischer:2022rxf,Aebischer:2023djt,Aebischer:2024xnf}.

%%%%%%%%%%%%%%%%%%%%%%%%%%%%%%%%%%%%%%%%%%%%%%%
\section*{Acknowledgments}
We thank Martin Gorbahn, Mikolaj Misiak, Jacky Kumar, Jason Aebischer and Marko Pesut for useful discussions. We thank Andrzej Buras, Mikolaj Misiak, Jason Aebisher and Marko Pesut for comments on the manuscript.
We are particularly grateful to Mikolaj Misiak for checking and confirming our final results.

P.M. acknowledges funding from the Spanish MCIN/AEI/10.13039/501100011033:
grant PRE2022-103999 funded by MCIN/AEI/10.13039/501100011033 and by "ESF Investing in your future", 
grant CEX2019-000918-M through the “Unit of Excellence Mar\'ia de Maeztu 2020-2023” award to the Institute of Cosmos Sciences. 

J.V. acknowledges funding from grant 2021-SGR-249 (Generalitat de Catalunya), and from the Spanish
MCIN/AEI/10.13039/501100011033 through the following grants: 
grant CNS2022-135262 funded by the “European Union NextGenerationEU/PRTR”,
grant RYC-2017-21870 funded by “ESF Investing in your future” through the ``Ram\'on y Cajal'' program,
grant CEX2019-000918-M through the “Unit of Excellence Mar\'ia de Maeztu 2020-2023” award to the Institute of Cosmos Sciences,
and grants PID2019-105614GB-C21 and PID2022-136224NB-C21.

%%%%%%%%%%%%%%%%%%%%%%%%%%%%%%%%%%%%%%%%%%%%%%%%%%%%%%%%%%%%%%%%%%%%%%%%%%%%%%%%%%%%

\appendix

\newpage
%%%%%%%%%%%%%%%%%%%%%%%%%%%%%%%%%%%%%%%%%%%%%%%
\section{BMU Operator Basis}
\label{app:OPBasis}

The physical operator basis we use and refer to throughout the text is the so-called BMU basis \cite{Buras:2000if} for $(\bar s d)(\bar q q)$ operators. The first two operators in this basis are the $u$-type
\begin{equation}
\begin{aligned}
    &Q_1 = (\bar{s}^{\alpha} \gamma^{\mu}P_L u^{\beta})(\bar{u}^{\beta} \gamma_{\mu}P_L d^{\alpha}) \;, \qquad 
    &Q_2 = (\bar{s}^{\alpha} \gamma^{\mu}P_L u^{\alpha})(\bar{u}^{\beta} \gamma_{\mu}P_L d^{\beta}) \;,
\end{aligned}
\end{equation}
where $\alpha,\beta$ are SU($N_c$) indices. We use also the alternative Fierz-transformed version of these two operators, also featured in \cite{Buras:2000if},
\begin{equation}
\begin{aligned}
    &\widetilde{Q}_1 = (\bar{s}^{\alpha} \gamma^{\mu}P_L d^{\alpha})(\bar{u}^{\beta} \gamma_{\mu}P_L u^{\beta}) \;, \qquad &\widetilde{Q}_2 = (\bar{s}^{\alpha} \gamma^{\mu}P_L d^{\beta})(\bar{u}^{\beta} \gamma_{\mu}P_L u^{\alpha}) \;.
\end{aligned}
\end{equation}
Following up, one has the four QCD penguin operators, summing over all flavors,
\begin{equation}
\begin{aligned}
    &Q_3 = (\bar{s}^{\alpha} \gamma^{\mu}P_L d^{\alpha})\sum_q(\bar{q}^{\beta} \gamma_{\mu}P_L q^{\beta}) \;, \qquad 
    &Q_4 = (\bar{s}^{\alpha} \gamma^{\mu}P_L d^{\beta})\sum_q(\bar{q}^{\beta} \gamma_{\mu}P_L q^{\alpha}) \;, \\
    &Q_5 = (\bar{s}^{\alpha} \gamma^{\mu}P_L d^{\alpha})\sum_q(\bar{q}^{\beta} \gamma_{\mu}P_R q^{\beta}) \;, \qquad 
    &Q_6 = (\bar{s}^{\alpha} \gamma^{\mu}P_L d^{\beta})\sum_q(\bar{q}^{\beta} \gamma_{\mu}P_R q^{\alpha}) \;,
\end{aligned}
\end{equation}
and the four QED penguins, again featuring a sum over flavors,
\begin{equation}
\begin{aligned}
    &Q_7 = \frac{3}{2}(\bar{s}^{\alpha} \gamma^{\mu}P_L d^{\alpha})\sum_q Q_q(\bar{q}^{\beta} \gamma_{\mu}P_R q^{\beta}) \;, \qquad 
    &Q_8 = \frac{3}{2}(\bar{s}^{\alpha} \gamma^{\mu}P_L d^{\beta})\sum_q Q_q(\bar{q}^{\beta} \gamma_{\mu}P_R q^{\alpha}) \;, \\
    &Q_9 = \frac{3}{2}(\bar{s}^{\alpha} \gamma^{\mu}P_L d^{\alpha})\sum_q Q_q(\bar{q}^{\beta} \gamma_{\mu}P_L q^{\beta}) \;, \qquad 
    &Q_{10} = \frac{3}{2}(\bar{s}^{\alpha} \gamma^{\mu}P_L d^{\beta})\sum_q Q_q(\bar{q}^{\beta} \gamma_{\mu}P_L q^{\alpha}) \;.
\end{aligned}
\end{equation}
These ten operators form the Standard Model sector, which is addressed in \cite{Buras:1989xd, Buras:1992tc}. The BMU basis then follows with a set of BSM operators, as introduced in \cite{Buras:2000if}, which starts with
\begin{equation}
\begin{aligned}
    &Q_{11} = (\bar{s}^{\alpha} \gamma^{\mu}P_L d^{\alpha}) (\bar{d}^{\beta} \gamma_{\mu}P_L d^{\beta}) + (\bar{s}^{\alpha} \gamma^{\mu}P_L d^{\alpha}) (\bar{s}^{\beta} \gamma_{\mu}P_L s^{\beta}) \;, \\
    &Q_{12} = (\bar{s}^{\alpha} \gamma^{\mu}P_L d^{\beta}) (\bar{d}^{\beta} \gamma_{\mu}P_R d^{\alpha}) + (\bar{s}^{\alpha} \gamma^{\mu}P_L d^{\beta}) (\bar{s}^{\beta} \gamma_{\mu}P_R s^{\alpha}) \;, \\ 
    &Q_{13} = (\bar{s}^{\alpha} \gamma^{\mu}P_L d^{\alpha}) (\bar{d}^{\beta} \gamma_{\mu}P_R d^{\beta}) + (\bar{s}^{\alpha} \gamma^{\mu}P_L d^{\alpha}) (\bar{s}^{\beta} \gamma_{\mu}P_R s^{\beta}) \;.
\end{aligned}
\end{equation}
In our discussion, we need only operators up to $Q_{11}$; although its Fierz-transformed version $Q'_{11}$ is also featured in the composition of the alternative operator $Q_{11}^+$ before Eq.~(\ref{VSLL+VXLL}),
\begin{equation}
    Q'_{11} = (\bar{s}^{\alpha} \gamma^{\mu}P_L d^{\beta}) (\bar{d}^{\beta} \gamma_{\mu}P_L d^{\alpha}) + (\bar{s}^{\alpha} \gamma^{\mu}P_L d^{\beta}) (\bar{s}^{\beta} \gamma_{\mu}P_L s^{\alpha}) \;.
\end{equation}
In addition, to refer to specific structures within operators, as in Eqs.~(\ref{VSLL}) and (\ref{VXLL}) we use the following general notation:
\begin{equation}
\begin{aligned}
    &\mathcal{O}^{VS,LL}_{ijkl} = (\bar{q}_i^{\alpha} \gamma^{\mu}P_L q_j^{\alpha})(\bar{q}_k^{\beta} \gamma_{\mu}P_L q_l^{\beta}) \;, \qquad &\mathcal{O}^{VX,LL}_{ijkl}  = (\bar{q}_i^{\alpha} \gamma^{\mu}P_L q_j^{\beta})(\bar{q}_k^{\beta} \gamma_{\mu}P_L q_l^{\alpha}) \;.
\end{aligned}
\end{equation}
Beyond the discussion given in this work, there are three more $d$-type BSM vector operators,
\begin{equation}
\begin{aligned}
    &Q_{14} = (\bar{s}^{\alpha} \gamma^{\mu}P_L d^{\alpha}) (\bar{d}^{\beta} \gamma_{\mu}P_L d^{\beta}) - (\bar{s}^{\alpha} \gamma^{\mu}P_L d^{\alpha}) (\bar{s}^{\beta} \gamma_{\mu}P_L s^{\beta}) \;, \\
    &Q_{15} = (\bar{s}^{\alpha} \gamma^{\mu}P_L d^{\beta}) (\bar{d}^{\beta} \gamma_{\mu}P_R d^{\alpha}) - (\bar{s}^{\alpha} \gamma^{\mu}P_L d^{\beta}) (\bar{s}^{\beta} \gamma_{\mu}P_R s^{\alpha}) \;, \\ 
    &Q_{16} = (\bar{s}^{\alpha} \gamma^{\mu}P_L d^{\alpha}) (\bar{d}^{\beta} \gamma_{\mu}P_R d^{\beta}) - (\bar{s}^{\alpha} \gamma^{\mu}P_L d^{\alpha}) (\bar{s}^{\beta} \gamma_{\mu}P_R s^{\beta}) \;,
\end{aligned}
\end{equation}
and two additional vector operators involving flavors $u$ and $c$,
\begin{equation}
\begin{aligned}
    &Q_{17} = (\bar{s}^{\alpha} \gamma^{\mu}P_L d^{\beta}) (\bar{u}^{\beta} \gamma_{\mu}P_R u^{\alpha}) - (\bar{s}^{\alpha} \gamma^{\mu}P_L d^{\beta}) (\bar{c}^{\beta} \gamma_{\mu}P_R c^{\alpha}) \;, \\
    &Q_{18} = (\bar{s}^{\alpha} \gamma^{\mu}P_L d^{\alpha}) (\bar{u}^{\beta} \gamma_{\mu}P_R u^{\beta}) - (\bar{s}^{\alpha} \gamma^{\mu}P_L d^{\alpha}) (\bar{c}^{\beta} \gamma_{\mu}P_R c^{\beta}) \;.
\end{aligned}
\end{equation}
The rest of operators are scalar. They can be divided in 6 chirality-mixed operators,
\begin{equation}
\begin{aligned}
\label{Q_SRL}
    &Q_{19} = (\bar{s}^{\alpha} P_R d^{\beta}) (\bar{u}^{\beta} P_L u^{\alpha}) \;, \quad &
    &Q_{20} = (\bar{s}^{\alpha} P_R d^{\alpha}) (\bar{u}^{\beta} P_L u^{\beta}) \;, \\ 
    &Q_{21} = (\bar{s}^{\alpha} P_R d^{\beta}) (\bar{c}^{\beta} P_L c^{\alpha}) \;, \quad &
    &Q_{22} = (\bar{s}^{\alpha} P_R d^{\alpha}) (\bar{c}^{\beta} P_L c^{\beta}) \;, \\ 
    &Q_{23} = (\bar{s}^{\alpha} P_R d^{\beta}) (\bar{b}^{\beta} P_L b^{\alpha}) \;, \quad &
    &Q_{24} = (\bar{s}^{\alpha} P_R d^{\alpha}) (\bar{b}^{\beta} P_L b^{\beta}) \;, \\ 
\end{aligned}
\end{equation}
and 16 scalar right-handed operators,
\begin{equation}
\begin{aligned}
\label{Q_SRR}
    &Q_{25} = (\bar{s}^{\alpha} P_R d^{\beta}) (\bar{d}^{\beta} P_R d^{\alpha}) \;, \quad &
    &Q_{26} = (\bar{s}^{\alpha} \sigma^{\mu\nu} P_R d^{\alpha}) (\bar{d}^{\beta} \sigma_{\mu\nu} P_R d^{\beta}) \;, \\ 
    &Q_{27} = (\bar{s}^{\alpha} P_R d^{\beta}) (\bar{s}^{\beta} P_R s^{\alpha}) \;, \quad &
    &Q_{28} = (\bar{s}^{\alpha} \sigma^{\mu\nu} P_R d^{\alpha}) (\bar{s}^{\beta} \sigma_{\mu\nu} P_R s^{\beta}) \;, \\
    &Q_{29} = (\bar{s}^{\alpha} P_R d^{\beta}) (\bar{u}^{\beta} P_R u^{\alpha}) \;, \quad &
    &Q_{30} = (\bar{s}^{\alpha} P_R d^{\alpha}) (\bar{u}^{\beta} P_R u^{\beta}) \;, \\ 
    &Q_{31} = (\bar{s}^{\alpha} \sigma^{\mu\nu} P_R d^{\beta}) (\bar{u}^{\beta} \sigma_{\mu\nu} P_R u^{\alpha}) \;, \quad &
    &Q_{32} = (\bar{s}^{\alpha} \sigma^{\mu\nu} P_R d^{\alpha}) (\bar{u}^{\beta} \sigma_{\mu\nu} P_R u^{\beta}) \;, \\ 
    &Q_{33} = (\bar{s}^{\alpha} P_R d^{\beta}) (\bar{c}^{\beta} P_R c^{\alpha}) \;, \quad &
    &Q_{34} = (\bar{s}^{\alpha} P_R d^{\alpha}) (\bar{c}^{\beta} P_R c^{\beta}) \;, \\ 
    &Q_{35} = (\bar{s}^{\alpha} \sigma^{\mu\nu} P_R d^{\beta}) (\bar{c}^{\beta} \sigma_{\mu\nu} P_R c^{\alpha}) \;, \quad &
    &Q_{36} = (\bar{s}^{\alpha} \sigma^{\mu\nu} P_R d^{\alpha}) (\bar{c}^{\beta} \sigma_{\mu\nu} P_R c^{\beta}) \;, \\ 
    &Q_{37} = (\bar{s}^{\alpha} P_R d^{\beta}) (\bar{b}^{\beta} P_R b^{\alpha}) \;, \quad &
    &Q_{38} = (\bar{s}^{\alpha} P_R d^{\alpha}) (\bar{b}^{\beta} P_R b^{\beta}) \;, \\ 
    &Q_{39} = (\bar{s}^{\alpha}     \sigma^{\mu\nu} P_R d^{\beta}) (\bar{b}^{\beta} \sigma_{\mu\nu} P_R b^{\alpha}) \;, \quad &
    &Q_{40} = (\bar{s}^{\alpha} \sigma^{\mu\nu} P_R d^{\alpha}) (\bar{b}^{\beta} \sigma_{\mu\nu} P_R b^{\beta}) \;. \\ 
\end{aligned}
\end{equation}
Many of the operators in this basis can be separated in blocks not connected by the RGE, as it can be seen in the block diagonal ADM in \App{app:BMU_ADM}. Apart from these 40 operators, there is an additional RGE-disconnected block of the same size corresponding to the opposite-chirality operators. 

This operator basis contains five quark flavors, corresponding to an EFT where the top quark has been integrated out. The bases for EFTs with lower numbers of active flavors (i.e. integrating out the bottom, the charm, etc.) can be readily obtained by eliminating some of the operators in the five-flavor EFT. For instance, a candidate for the four-flavor ($f=4$) basis, corresponding to integrating out the $b$-quark, is obtained by eliminating the four QED penguins ($Q_{7}-Q_{10}$) and all scalar operators containing $b$-quarks ($Q_{23},Q_{24},Q_{37}-Q_{40}$). A three-flavor ($f=3$) basis, can then be obtained by eliminating also $Q_1,Q_2,Q_{17}$,$Q_{18}$, and all scalar operators with $c$-quarks ($Q_{21},Q_{22},Q_{33}-Q_{36}$).

%%%%%%%%%%%%%%%%%%%%%%%%%%%%%%%%%%%%%%%%%%%%%%%
\section{Evanescent Operator Basis}
\label{app:EVBasis}

The set of evanescent operators we use to specify the renormalization scheme for the two-loop ADM in the case of $\bar{s}b\bar{q}q$ operators is analogous to the ones given in~\Reff{Buras:2000if} for sectors $\bar{s}d\bar{u}c$ and $\bar{s}d\bar{s}d$, equivalent to the choice $a_{\text{ev}},b_{\text{ev}},c_{\text{ev}}, ... = 1$ in \Reff{Dekens:2019ept}. We list them here separated for generic flavors $q = u,c,d,s,b$, noting that they become redundant for $q = d,s$. In such case the \textit{tilde} evanescents are absent (that is, $\widetilde{E}^{\text{X}(q)}_i$ exist only for $q=u,c,b$). 

An evanescent basis defined in this manner, with analogous structures for each flavor (i.e. ensuring the same $d$-dimensional Fierz identities for all flavors), satisfies the condition discussed above \Eq{E_11}.%, necessary to observe flavor symmetries in the ADM.

Let us also note that any linear rotation of this evanescent basis ($E_i' = W_{ij}E_j$), involving no physical operators, leaves the physical anomalous dimensions unaltered. Therefore, any such evanescent basis defines a completely equivalent renormalization scheme.

Again, we limit our exposition to half of the total basis, given that the definition of the chiral-opposite sector is straightforward, $P_L \leftrightarrow P_R$. Starting with the VLL sector,
\allowdisplaybreaks
\eqa{
E^{\text{VLL}(q)}_1 &=& (\bar{s}^{\alpha}\gamma^{\mu}P_Lq^{\beta})(\bar{q}^{\beta}\gamma_{\mu}P_Ld^{\alpha}) - (\bar{s}^{\alpha}\gamma^{\mu}P_Ld^{\alpha})(\bar{q}^{\beta}\gamma_{\mu}P_Lq^{\beta}) \;, 
\nonumber\\
\widetilde{E}^{\text{VLL}(q)}_1 &=& (\bar{s}^{\alpha}\gamma^{\mu}P_Lq^{\alpha})(\bar{q}^{\beta}\gamma_{\mu}P_Ld^{\beta}) - (\bar{s}^{\alpha}\gamma^{\mu}P_Ld^{\beta})(\bar{q}^{\beta}\gamma_{\mu}P_Lq^{\alpha}) \;,
\nonumber\\
E^{\text{VLL}(q)}_2 &=& (\bar{s}^{\alpha}\gamma^{\mu}\gamma^{\nu}\gamma^{\rho}P_Ld^{\alpha})(\bar{q}^{\beta}\gamma_{\mu}\gamma_{\nu}\gamma_{\rho}P_Lq^{\beta}) - (16-4\epsilon)(\bar{s}^{\alpha}\gamma^{\mu}P_Ld^{\alpha})(\bar{q}^{\beta}\gamma_{\mu}P_Lq^{\beta}) \;, 
\nonumber\\
E^{\text{VLL}(q)}_3 &=& (\bar{s}^{\alpha}\gamma^{\mu}\gamma^{\nu}\gamma^{\rho}P_Ld^{\beta})(\bar{q}^{\beta}\gamma_{\mu}\gamma_{\nu}\gamma_{\rho}P_Lq^{\alpha}) - (16-4\epsilon)(\bar{s}^{\alpha}\gamma^{\mu}P_Ld^{\beta})(\bar{q}^{\beta}\gamma_{\mu}P_Lq^{\alpha}) \;, 
\\
\widetilde{E}^{\text{VLL}(q)}_2 &=& (\bar{s}^{\alpha}\gamma^{\mu}\gamma^{\nu}\gamma^{\rho}P_Lq^{\alpha})(\bar{q}^{\beta}\gamma_{\mu}\gamma_{\nu}\gamma_{\rho}P_Ld^{\beta}) - (16-4\epsilon)(\bar{s}^{\alpha}\gamma^{\mu}P_Lq^{\alpha})(\bar{q}^{\beta}\gamma_{\mu}P_Ld^{\beta}) \;, 
\nonumber\\
\widetilde{E}^{\text{VLL}(q)}_3 &=& (\bar{s}^{\alpha}\gamma^{\mu}\gamma^{\nu}\gamma^{\rho}P_Lq^{\beta})(\bar{q}^{\beta}\gamma_{\mu}\gamma_{\nu}\gamma_{\rho}P_Ld^{\alpha}) - (16-4\epsilon)(\bar{s}^{\alpha}\gamma^{\mu}P_Lq^{\beta})(\bar{q}^{\beta}\gamma_{\mu}P_Ld^{\alpha}) \;.
\nonumber
}
As for the VLR sector,
\eqa{
\allowdisplaybreaks
E^{\text{VLR}(q)}_1 &=&
2(\bar{s}^{\alpha}P_Rq^{\beta})(\bar{q}^{\beta}P_Ld^{\alpha}) - (\bar{s}^{\alpha}\gamma^{\mu}P_Ld^{\alpha})(\bar{q}^{\beta}\gamma_{\mu}P_Rq^{\beta}) \;, 
\nonumber\\
E^{\text{VLR}(q)}_2 &=&
2(\bar{s}^{\alpha}P_Rq^{\alpha})(\bar{q}^{\beta}P_Ld^{\beta}) - (\bar{s}^{\alpha}\gamma^{\mu}P_Ld^{\beta})(\bar{q}^{\beta}\gamma_{\mu}P_Rq^{\alpha}) \;, 
\nonumber\\
\widetilde{E}^{\text{VLR}(q)}_1 &=& 2(\bar{s}^{\alpha}P_Rd^{\beta})(\bar{q}^{\beta}P_Lq^{\alpha}) - (\bar{s}^{\alpha}\gamma^{\mu}P_Lq^{\alpha})(\bar{q}^{\beta}\gamma_{\mu}P_Rd^{\beta}) \;, 
\nonumber\\
\widetilde{E}^{\text{VLR}(q)}_2 &=& 2(\bar{s}^{\alpha}P_Rd^{\alpha})(\bar{q}^{\beta}P_Lq^{\beta}) - (\bar{s}^{\alpha}\gamma^{\mu}P_Lq^{\beta})(\bar{q}^{\beta}\gamma_{\mu}P_Rd^{\alpha}) \;, 
\\
E^{\text{VLR}(q)}_3 &=& (\bar{s}^{\alpha}\gamma^{\mu}\gamma^{\nu}\gamma^{\rho}P_Ld^{\alpha})(\bar{q}^{\beta}\gamma_{\mu}\gamma_{\nu}\gamma_{\rho}P_Rq^{\beta}) - (4+4\epsilon)(\bar{s}^{\alpha}\gamma^{\mu}P_Ld^{\alpha})(\bar{q}^{\beta}\gamma_{\mu}P_Rq^{\beta}) \;, 
\nonumber\\
E^{\text{VLR}(q)}_4 &=& (\bar{s}^{\alpha}\gamma^{\mu}\gamma^{\nu}\gamma^{\rho}P_Ld^{\beta})(\bar{q}^{\beta}\gamma_{\mu}\gamma_{\nu}\gamma_{\rho}P_Rq^{\alpha}) - (4+4\epsilon)(\bar{s}^{\alpha}\gamma^{\mu}P_Ld^{\beta})(\bar{q}^{\beta}\gamma_{\mu}P_Rq^{\alpha}) \;, 
\nonumber\\
\widetilde{E}^{\text{VLR}(q)}_3 &=& (\bar{s}^{\alpha}\gamma^{\mu}\gamma^{\nu}\gamma^{\rho}P_Lq^{\alpha})(\bar{q}^{\beta}\gamma_{\mu}\gamma_{\nu}\gamma_{\rho}P_Rd^{\beta}) - (4+4\epsilon)(\bar{s}^{\alpha}\gamma^{\mu}P_Lq^{\alpha})(\bar{q}^{\beta}\gamma_{\mu}P_Rd^{\beta}) \;, 
\nonumber\\
\widetilde{E}^{\text{VLR}(q)}_4 &=& (\bar{s}^{\alpha}\gamma^{\mu}\gamma^{\nu}\gamma^{\rho}P_Lq^{\beta})(\bar{q}^{\beta}\gamma_{\mu}\gamma_{\nu}\gamma_{\rho}P_Rd^{\alpha}) - (4+4\epsilon)(\bar{s}^{\alpha}\gamma^{\mu}P_Lq^{\beta})(\bar{q}^{\beta}\gamma_{\mu}P_Rd^{\alpha}) \;.
\nonumber
}
For the SRL sector, Fierz-related to VLR,
\eqa{
E^{\text{SRL}(q)}_1 &=& (\bar{s}^{\alpha}\sigma^{\mu\nu}P_Rd^{\alpha})(\bar{q}^{\beta}\sigma_{\mu\nu}P_Lq^{\beta}) - 6\epsilon(\bar{s}^{\alpha}P_Rd^{\alpha})(\bar{q}^{\beta}P_Lq^{\beta}) \;, 
\nonumber\\
E^{\text{SRL}(q)}_2 &=& (\bar{s}^{\alpha}\sigma^{\mu\nu}P_Rd^{\beta})(\bar{q}^{\beta}\sigma_{\mu\nu}P_Lq^{\alpha}) - 6\epsilon(\bar{s}^{\alpha}P_Rd^{\beta})(\bar{q}^{\beta}P_Lq^{\alpha}) \;, 
\nonumber\\
\widetilde{E}^{\text{SRL}(q)}_1 &=& (\bar{s}^{\alpha}\sigma^{\mu\nu}P_Rq^{\alpha})(\bar{q}^{\beta}\sigma_{\mu\nu}P_Ld^{\beta}) - 6\epsilon(\bar{s}^{\alpha}P_Rq^{\alpha})(\bar{q}^{\beta}P_Ld^{\beta}) \;, 
\\
\widetilde{E}^{\text{SRL}(q)}_2 &=& (\bar{s}^{\alpha}\sigma^{\mu\nu}P_Rq^{\beta})(\bar{q}^{\beta}\sigma_{\mu\nu}P_Ld^{\alpha}) - 6\epsilon(\bar{s}^{\alpha}P_Rq^{\beta})(\bar{q}^{\beta}P_Ld^{\alpha}) \;.
\nonumber
}
Finally, for the SRR sector,
\eqa{
E^{\text{SRR}(q)}_1 &=& (\bar{s}^{\alpha}\sigma^{\mu\nu}P_Rd^{\alpha})(\bar{q}^{\beta}\sigma_{\mu\nu}P_Rq^{\beta}) + 4(\bar{s}^{\alpha}P_Rd^{\alpha})(\bar{q}^{\beta}P_Rq^{\beta}) + 8(\bar{s}^{\alpha}P_Rq^{\beta})(\bar{q}^{\beta}P_Rd^{\alpha}) \;, 
\nonumber\\
E^{\text{SRR}(q)}_2 &=& (\bar{s}^{\alpha}\sigma^{\mu\nu}P_Rd^{\beta})(\bar{q}^{\beta}\sigma_{\mu\nu}P_Rq^{\alpha}) + 4(\bar{s}^{\alpha}P_Rd^{\beta})(\bar{q}^{\beta}P_Rq^{\alpha}) + 8(\bar{s}^{\alpha}P_Rq^{\alpha})(\bar{q}^{\beta}P_Rd^{\beta}) \;, 
\nonumber\\
\widetilde{E}^{\text{SRR}(q)}_1 &=& (\bar{s}^{\alpha}\sigma^{\mu\nu}P_Rq^{\alpha})(\bar{q}^{\beta}\sigma_{\mu\nu}P_Rd^{\beta}) + 4(\bar{s}^{\alpha}P_Rq^{\alpha})(\bar{q}^{\beta}P_Rd^{\beta}) + 8(\bar{s}^{\alpha}P_Rd^{\beta})(\bar{q}^{\beta}P_Rq^{\alpha}) \;, 
\nonumber\\
\widetilde{E}^{\text{SRR}(q)}_2 &=& (\bar{s}^{\alpha}\sigma^{\mu\nu}P_Rq^{\beta})(\bar{q}^{\beta}\sigma_{\mu\nu}P_Rd^{\alpha}) + 4(\bar{s}^{\alpha}P_Rq^{\beta})(\bar{q}^{\beta}P_Rd^{\alpha}) + 8(\bar{s}^{\alpha}P_Rd^{\alpha})(\bar{q}^{\beta}P_Rq^{\beta}) \;, 
\nonumber\\
E^{\text{SRR}(q)}_3 &=& (\bar{s}^{\alpha}\gamma^{\mu}\gamma^{\nu}\gamma^{\rho}\gamma^{\sigma}P_Rd^{\alpha})(\bar{q}^{\beta}\gamma_{\mu}\gamma_{\nu}\gamma_{\rho}\gamma_{\sigma}P_Rq^{\beta}) - (64 - 96\epsilon) (\bar{s}^{\alpha}P_Rd^{\alpha})(\bar{q}^{\beta}P_Rq^{\beta}) 
\nonumber\\
&& + (16 - 8\epsilon)(\bar{s}^{\alpha}\sigma^{\mu\nu}P_Rd^{\alpha})(\bar{q}^{\beta}\sigma_{\mu\nu}P_Rq^{\beta}) \;, 
\\
E^{\text{SRR}(q)}_4 &=& (\bar{s}^{\alpha}\gamma^{\mu}\gamma^{\nu}\gamma^{\rho}\gamma^{\sigma}P_Rd^{\beta})(\bar{q}^{\beta}\gamma_{\mu}\gamma_{\nu}\gamma_{\rho}\gamma_{\sigma}P_Rq^{\alpha}) - (64 - 96\epsilon) (\bar{s}^{\alpha}P_Rd^{\beta})(\bar{q}^{\beta}P_Rq^{\alpha}) 
\nonumber\\
&& + (16 - 8\epsilon)(\bar{s}^{\alpha}\sigma^{\mu\nu}P_Rd^{\beta})(\bar{q}^{\beta}\sigma_{\mu\nu}P_Rq^{\alpha}) \;, 
\nonumber\\
\widetilde{E}^{\text{SRR}(q)}_3 &=& (\bar{s}^{\alpha}\gamma^{\mu}\gamma^{\nu}\gamma^{\rho}\gamma^{\sigma}P_Rq^{\alpha})(\bar{q}^{\beta}\gamma_{\mu}\gamma_{\nu}\gamma_{\rho}\gamma_{\sigma}P_Rd^{\beta}) - (64 - 96\epsilon) (\bar{s}^{\alpha}P_Rq^{\alpha})(\bar{q}^{\beta}P_Rd^{\beta}) 
\nonumber\\
&& + (16 - 8\epsilon)(\bar{s}^{\alpha}\sigma^{\mu\nu}P_Rq^{\alpha})(\bar{q}^{\beta}\sigma_{\mu\nu}P_Rd^{\beta}) \;, 
\nonumber\\
\widetilde{E}^{\text{SRR}(q)}_4 &=& (\bar{s}^{\alpha}\gamma^{\mu}\gamma^{\nu}\gamma^{\rho}\gamma^{\sigma}P_Rq^{\beta})(\bar{q}^{\beta}\gamma_{\mu}\gamma_{\nu}\gamma_{\rho}\gamma_{\sigma}P_Rd^{\alpha}) - (64 - 96\epsilon) (\bar{s}^{\alpha}P_Rq^{\beta})(\bar{q}^{\beta}P_Rd^{\alpha}) 
\nonumber\\
&& + (16 - 8\epsilon)(\bar{s}^{\alpha}\sigma^{\mu\nu}P_Rq^{\beta})(\bar{q}^{\beta}\sigma_{\mu\nu}P_Rd^{\alpha}) \;.
\nonumber
}

%%%%%%%%%%%%%%%%%%%%%%%%%%%%%%%%%%%%%%%%%%%%%%%
\section{Full Anomalous Dimension Matrix to NLO in QCD}
\label{app:BMU_ADM}

We provide here the complete one- and two-loop ADMs, for the BMU basis as presented in the previous appendix. These ADMs include current-current and penguin contributions, the latter accounting already for the correction discussed in this work (in {\color{red}{red}}). As in~\App{app:OPBasis}, we will limit ourselves to half of the basis, with the other half being its chiral-opposite, given that the full matrix corresponds to two identical copies of the one we shall provide here.

In these expressions, $f$ will be the number of active quark flavors, $u,d$ stand for the number of active up- and down-type quarks, respectively. The references to $f$ in these ADMs allow for the determination of the corresponding anomalous dimensions in theories with a different number of active quark flavors. Strictly speaking, the full set of matrices given in this appendix correspond to the five-flavor theory ($f=5$). Going to lower numbers of active flavors not only changes the value of $f$, but also requires for the elimination of all rows and columns corresponding to redundant operators ``integrated out'' from the basis, as explained in~\App{app:OPBasis}.

\subsection{Leading Order}

The LO ADM can be written in terms of two main blocks,
\eqa{
\label{eq:ADM0_block_I}
\hat{\gamma}^{(0)}_{\text{BMU}} &=&
\left(
\begin{array}{ccc}
\hat{\gamma}^{(0)}_{\text{VLV}} & 0 \\
0 & \hat{\gamma}^{(0)}_{\text{SRS}} \\
\end{array}
\right)
\ .
}
The first block corresponds to the 18 vector operators $\{Q_1{} - Q_{18} \}$, and thus contains all penguin contributions,
\eqa{
\label{eq:ADM0_VLV}
\hat{\gamma}^{(0)}_{\text{VLV}} &=
\left(
\begin{array}{ccccc}
\hat{\gamma}^{(0)}_{CC} & \hat{\gamma}^{(0)}_{CC\to P} & 0 & 0 & 0 \\
0 & \hat{\gamma}^{(0)}_P & 0 & 0 & 0 \\
0 & \hat{\gamma}^{(0)}_{d+s\to P} & \hat{\gamma}^{(0)}_{d+s} & 0 & 0 \\
0 & 0 & 0 & \hat{\gamma}^{(0)}_{d-s} & 0 \\
0 & 0 & 0 & 0 & \hat{\gamma}^{(0)}_{u-c} \\
\end{array}
\right)
\ ,
\qquad
\hat{\gamma}^{(0)}_P =&
\left(
\begin{array}{cc}
 \hat{\gamma}^{(0)}_{PP} & 0 \\
 \hat{\gamma}^{(0)}_{QP} & \hat{\gamma}^{(0)}_{QQ} \\
\end{array}
\right)
\ .
}
The other term in Eq.~({\ref{eq:ADM0_block_I}}) is block-diagonal, and involves the 22 scalar operators $\{Q_{19} - Q_{40} \}$,
\eqa{
\label{eq:ADM0_SRS}
\hat{\gamma}^{(0)}_{\text{SRS}} &= \text{diag}\Big(\hat{\gamma}^{(0)}_{\text{SRL}(u)},\hat{\gamma}^{(0)}_{\text{SRL}(c)},\hat{\gamma}^{(0)}_{\text{SRL}(b)},\hat{\gamma}^{(0)}_{\text{SRR}(d)},\hat{\gamma}^{(0)}_{\text{SRR}(s)},\hat{\gamma}^{(0)}_{\text{SRR}(u)},\hat{\gamma}^{(0)}_{\text{SRR}(c)},\hat{\gamma}^{(0)}_{\text{SRR}(b)}\Big)
\ .
}
The first three blocks here $\hat{\gamma}^{(0)}_{\text{SRL}(u,c,b)}$ are identical $2 \times 2$ matrices corresponding to the operators in~\Eq{Q_SRL}. The following two blocks $\hat{\gamma}^{(0)}_{\text{SRR}(d,s)}$ are again identical and $2 \times 2$, corresponding to the first four operators in~\Eq{Q_SRR}. The remaining three blocks $\hat{\gamma}^{(0)}_{\text{SRR}(u,c,b)}$ are identical $4 \times 4$ matrices, and correspond to the last twelve operators in~\Eq{Q_SRR}.

The individual blocks in Eqs.~(\ref{eq:ADM0_VLV}) and (\ref{eq:ADM0_SRS}) read, fixing the number of colors in the QCD gauge group SU($N_c$) to $N_c = 3$,
\begin{gather}
    \hat{\gamma}^{(0)}_{CC} =
    \left(
    \begin{array}{cc}
     -2 & 6 \\
     6 & -2 \\
    \end{array}
    \right)
    \ ,
    \qquad
    \hat{\gamma}^{(0)}_{CC \rightarrow P} =
    \left(
    \begin{array}{cccccccc}
    0 & 0 & 0 & 0 \\
     -\frac{2}{9} & \frac{2}{3} & -\frac{2}{9} & \frac{2}{3} \\
    \end{array}
    \right)
    \ ,
    \\
    \hat{\gamma}^{(0)}_{PP} =
    \left(
    \begin{array}{cccc}
     -\frac{22}{9} & \frac{22}{3} & -\frac{4}{9} & \frac{4}{3} \\
     6-\frac{2 f}{9} & \frac{2 f}{3}-2 & -\frac{2 f}{9} & \frac{2 f}{3} \\
     0 & 0 & 2 & -6 \\
     -\frac{2 f}{9} & \frac{2 f}{3} & -\frac{2 f}{9} & \frac{2 f}{3}-16 \\
    \end{array}
    \right)
    \ ,
    \qquad
    \hat{\gamma}^{(0)}_{QQ} =
    \left(
    \begin{array}{cccc}
     2 & -6 & 0 & 0 \\
     0 & -16 & 0 & 0 \\
     0 & 0 & -2 & 6 \\
     0 & 0 & 6 & -2 \\
    \end{array}
    \right)
    \ ,
    \\
    \hat{\gamma}^{(0)}_{QP} =
    \left(
    \begin{array}{cccc}
     0 & 0 & 0 & 0 \\
     -\frac{2(u-d/2)}{9}  & \frac{2(u-d/2)}{3} & -\frac{2(u-d/2)}{9}  & \frac{2(u-d/2)}{3} \\
     \frac{2}{9} & -\frac{2}{3} & \frac{2}{9} & -\frac{2}{3} \\
     -\frac{2(u-d/2)}{9}  & \frac{2(u-d/2)}{3} & -\frac{2(u-d/2)}{9}  & \frac{2(u-d/2)}{3} \\
    \end{array}
    \right)
    \ ,
    \qquad
    \hat{\gamma}^{(0)}_{d+s \rightarrow P} =
    \left(
    \begin{array}{cccc}
     -\frac{4}{9} & \frac{4}{3} & -\frac{4}{9} & \frac{4}{3} \\
     -\frac{4}{9} & \frac{4}{3} & -\frac{4}{9} & \frac{4}{3} \\
     0 & 0 & 0 & 0 \\
    \end{array}
    \right)
    \ ,
    \\
    \hat{\gamma}^{(0)}_{d+s} = \hat{\gamma}^{(0)}_{d-s} =
    \left(
    \begin{array}{ccc}
     4 & 0 & 0 \\
     0 & -16 & 0 \\
     0 & -6 & 2 \\
    \end{array}
    \right)
    \ ,
    \qquad
    \hat{\gamma}^{(0)}_{u-c} =
    \left(
    \begin{array}{cc}
     -16 & 0 \\
     -6 & 2 \\
    \end{array}
    \right)
    \ ,
    \\
    \hat{\gamma}^{(0)}_{\text{SRL}(u)} = \hat{\gamma}^{(0)}_{\text{SRL}(c)} = \hat{\gamma}^{(0)}_{\text{SRL}(b)} =
    \left(
    \begin{array}{cc}
     2 & -6 \\
     0 & -16 \\
    \end{array}
    \right)
    \ ,
    \qquad
    \hat{\gamma}^{(0)}_{\text{SRL}(d)} = \hat{\gamma}^{(0)}_{\text{SRL}(s)} = 
    \left(
    \begin{array}{cc}
     -10 & -\frac{1}{6} \\
     40 & \frac{34}{3} \\
    \end{array}
    \right)
    \ ,
    \\
    \hat{\gamma}^{(0)}_{\text{SRR}(u)} = \hat{\gamma}^{(0)}_{\text{SRR}(c)} = \hat{\gamma}^{(0)}_{\text{SRR}(b)} =
    \left(
    \begin{array}{cccc}
     2 & -6 & -\frac{7}{6} & -\frac{1}{2} \\
     0 & -16 & -1 & \frac{1}{3} \\
     -56 & -24 & -\frac{38}{3} & 6 \\
     -48 & 16 & 0 & \frac{16}{3} \\
    \end{array}
    \right)
    \ .
\end{gather}

\subsection{Next-to-Leading Order}

The NLO ADM can also be written in terms of two main blocks,
\eqa{
\label{eq:ADM1_block_I}
\hat{\gamma}^{(1)}_{\text{BMU}} &=&
\left(
\begin{array}{ccc}
\hat{\gamma}^{(1)}_{\text{VLV}} & 0 \\
0 & \hat{\gamma}^{(1)}_{\text{SRS}} \\
\end{array}
\right)
\ .
}
The first block corresponds to the 18 vector operators $\{Q_1{} - Q_{18} \}$, and thus contains all penguin contributions,
\eqa{
\label{eq:ADM1_VLV}
\hat{\gamma}^{(1)}_{\text{VLV}} &=
\left(
\begin{array}{ccccc}
\hat{\gamma}^{(1)}_{CC} & \hat{\gamma}^{(1)}_{CC\to P} & 0 & 0 & 0 \\
0 & \hat{\gamma}^{(1)}_P & 0 & 0 & 0 \\
0 & {\color{red}\hat{\gamma}^{(1)}_{d+s\to P}} & \hat{\gamma}^{(1)}_{d+s} & 0 & 0 \\
0 & 0 & 0 & \hat{\gamma}^{(1)}_{d-s} & 0 \\
0 & 0 & 0 & 0 & \hat{\gamma}^{(1)}_{u-c} \\
\end{array}
\right)
\ ,
\qquad
\hat{\gamma}^{(1)}_P =&
\left(
\begin{array}{cc}
 \hat{\gamma}^{(1)}_{PP} & 0 \\
 \hat{\gamma}^{(1)}_{QP} & \hat{\gamma}^{(1)}_{QQ} \\
\end{array}
\right)
\ .
}
The other term in Eq.~({\ref{eq:ADM1_block_I}}) is block-diagonal, and involves the 22 scalar operators $\{Q_{19} - Q_{40} \}$,
\eqa{
\label{eq:ADM1_SRS}
\hat{\gamma}^{(1)}_{\text{SRS}} &= \text{diag}\Big(\hat{\gamma}^{(1)}_{\text{SRL}(u)},\hat{\gamma}^{(1)}_{\text{SRL}(c)},\hat{\gamma}^{(1)}_{\text{SRL}(b)},\hat{\gamma}^{(1)}_{\text{SRR}(d)},\hat{\gamma}^{(1)}_{\text{SRR}(s)},\hat{\gamma}^{(1)}_{\text{SRR}(u)},\hat{\gamma}^{(1)}_{\text{SRR}(c)},\hat{\gamma}^{(1)}_{\text{SRR}(b)}\Big)
\ .
}
The correspondence to the respective operators is analogous to the one in Eq.~(\ref{eq:ADM0_SRS}).

The individual blocks in Eqs.~(\ref{eq:ADM1_VLV}) and (\ref{eq:ADM1_SRS}) read, fixing the number of colors in the QCD gauge group SU($N_c$) to $N_c = 3$,
\begin{gather}
    \hat{\gamma}^{(1)}_{CC} =
    \left(
    \begin{array}{cc}
     -\frac{2 f}{9}-\frac{21}{2} & \frac{2 f}{3}+\frac{7}{2} \\
     \frac{2 f}{3}+\frac{7}{2} & -\frac{2 f}{9}-\frac{21}{2} \\
    \end{array}
    \right)
    \ ,
    \qquad
    \hat{\gamma}^{(1)}_{CC \rightarrow P} =
    \left(
    \begin{array}{cccccccc}
     \frac{79}{9} & -\frac{7}{3} & -\frac{65}{9} & -\frac{7}{3} & 0 & 0 & 0 & 0 \\
     -\frac{202}{243} & \frac{1354}{81} & -\frac{1192}{243} & \frac{904}{81} & 0 & 0 & 0 & 0 \\
    \end{array}
    \right)
    \ ,
    \\
    \hat{\gamma}^{(1)}_{PP} =
    \left(
    \begin{array}{cccc}
     \frac{71 f}{9}-\frac{5911}{486} & \frac{f}{3}+\frac{5983}{162} & -\frac{71 f}{9}-\frac{2384}{243} & \frac{1808}{81}-\frac{f}{3} \\
     \frac{56 f}{243}+\frac{379}{18} & \frac{808 f}{81}-\frac{91}{6} & -\frac{502 f}{243}-\frac{130}{9} & \frac{646 f}{81}-\frac{14}{3} \\
     -\frac{61 f}{9} & -\frac{11 f}{3} & \frac{61 f}{9}+\frac{71}{3} & \frac{11 f}{3}-99 \\
     -\frac{682 f}{243} & \frac{106 f}{81} & \frac{1676 f}{243}-\frac{225}{2} & \frac{1348 f}{81}-\frac{1343}{6} \\
    \end{array}
    \right)
    \ ,
    \\
    \hat{\gamma}^{(1)}_{QP} =
    \left(
    \begin{array}{cccc}
     \frac{61 d}{18}-\frac{61 u}{9} & \frac{11 d}{6}-\frac{11 u}{3} & \frac{83 u}{9}-\frac{83 d}{18} & \frac{11 d}{6}-\frac{11 u}{3} \\
     \frac{341 d}{243}-\frac{682 u}{243} & \frac{106 u}{81}-\frac{53 d}{81} & \frac{704 u}{243}-\frac{352 d}{243} & \frac{736 u}{81}-\frac{368 d}{81} \\
     -\frac{73 d}{18}+\frac{73 u}{9}+\frac{202}{243} & \frac{d}{6}-\frac{u}{3}-\frac{1354}{81} & \frac{71 d}{18}-\frac{71 u}{9}+\frac{1192}{243} & \frac{d}{6}-\frac{u}{3}-\frac{904}{81} \\
     \frac{53 d}{243}-\frac{106 u}{243}-\frac{79}{9} & -\frac{413 d}{81}+\frac{826 u}{81}+\frac{7}{3} & \frac{251 d}{243}-\frac{502 u}{243}+\frac{65}{9} & -\frac{323 d}{81}+\frac{646 u}{81}+\frac{7}{3} \\
    \end{array}
    \right)
    \ ,
    \\
    \hat{\gamma}^{(1)}_{QQ} =
    \left(
    \begin{array}{cccc}
     \frac{71}{3}-\frac{22 f}{9} & \frac{22 f}{3}-99 & 0 & 0 \\
     4 f-\frac{225}{2} & \frac{68 f}{9}-\frac{1343}{6} & 0 & 0 \\
     0 & 0 & -\frac{2 f}{9}-\frac{21}{2} & \frac{2 f}{3}+\frac{7}{2} \\
     0 & 0 & \frac{2 f}{3}+\frac{7}{2} & -\frac{2 f}{9}-\frac{21}{2} \\
    \end{array}
    \right)
    \ ,
    \\
    {\color{red}\hat{\gamma}^{(1)}_{d+s \rightarrow P}} =
    \left(
    \begin{array}{cccccccc}
     \color{red}{\frac{3862}{243}} & \color{red}{\frac{2330}{81}} & \color{red}{-\frac{5894}{243}} & \color{red}{\frac{1430}{81}} & 0 & 0 & 0 & 0 \\
     -\frac{1364}{243} & \frac{212}{81} & \frac{1408}{243} & \frac{1472}{81} & 0 & 0 & 0 & 0 \\
     -\frac{122}{9} & -\frac{22}{3} & \frac{166}{9} & -\frac{22}{3} & 0 & 0 & 0 & 0 \\
    \end{array}
    \right)
    \ ,
    \\
    \hat{\gamma}^{(1)}_{d+s} = \hat{\gamma}^{(1)}_{d-s} =
    \left(
    \begin{array}{ccc}
     \frac{4 f}{9}-7 & 0 & 0 \\
     0 & \frac{68 f}{9}-\frac{1343}{6} & 4 f-\frac{225}{2} \\
     0 & \frac{22 f}{3}-99 & \frac{71}{3}-\frac{22 f}{9} \\
    \end{array}
    \right)
    \ ,
    \qquad
    \hat{\gamma}^{(1)}_{u-c} =
    \left(
    \begin{array}{cc}
     \frac{68 f}{9}-\frac{1343}{6} & 4 f-\frac{225}{2} \\
     \frac{22 f}{3}-99 & \frac{71}{3}-\frac{22 f}{9} \\
    \end{array}
    \right)
    \ ,
    \\
    \hat{\gamma}^{(1)}_{\text{SRL}(u)} = \hat{\gamma}^{(1)}_{\text{SRL}(c)} = \hat{\gamma}^{(1)}_{\text{SRL}(b)} =
    \left(
    \begin{array}{cc}
     \frac{71}{3}-\frac{22 f}{9} & \frac{22 f}{3}-99 \\
     4 f-\frac{225}{2} & \frac{68 f}{9}-\frac{1343}{6} \\
    \end{array}
    \right)
    \ ,
    \\
    \hat{\gamma}^{(1)}_{\text{SRL}(d)} = \hat{\gamma}^{(1)}_{\text{SRL}(s)} = 
    \left(
    \begin{array}{cc}
     \frac{74 f}{9}-\frac{1459}{9} & \frac{f}{54}+\frac{35}{36} \\
     \frac{6332}{9}-\frac{584 f}{9} & \frac{2065}{9}-\frac{394 f}{27} \\
    \end{array}
    \right)
    \ ,
    \\
    \hat{\gamma}^{(1)}_{\text{SRR}(u)} = \hat{\gamma}^{(1)}_{\text{SRR}(c)} = \hat{\gamma}^{(1)}_{\text{SRR}(b)} =
    \left(
    \begin{array}{cccc}
     \frac{350}{9}-\frac{64 f}{9} & \frac{16 f}{3}-\frac{470}{3} & \frac{7 f}{54}-\frac{805}{36} & \frac{f}{18}+\frac{77}{12} \\
     -\frac{130}{3} & \frac{80 f}{9}-\frac{2710}{9} & \frac{f}{9}-\frac{31}{2} & \frac{61}{18}-\frac{f}{27} \\
     \frac{616 f}{9}-\frac{12292}{9} & \frac{88 f}{3}-\frac{2908}{3} & \frac{200 f}{27}-\frac{1262}{9} & 50-\frac{8 f}{3} \\
     \frac{176 f}{3}-\frac{1880}{3} & \frac{2648}{9}-\frac{176 f}{9} & \frac{8 f}{3}+\frac{26}{3} & \frac{1582}{9}-\frac{232 f}{27} \\
    \end{array}
    \right)
    \ .
\end{gather}
Again, we have indicated in \red{red} the entries that are different from BMU.

%%%%%%%%%%%%%%%%%%%%%%%%%%%%%%%%%%%%%%%%%%%%%%%%%%%%%%%%%%%%%%%%%%%%%%%%%%%%%%%%%%%%

%%%%%%%%%%%%%%%%%%%%%%%%%%%%%%%%%%%%%%%%%%%%%%%%%%%%%%%%%%%%%%%%%%%%%%%%%%%%%%%%%%%%
%%%%%%%%%%%%%%%%%%%%%%%%%%%%%%%%%%%%%%%%%%%%%%%%%%%%%%%%%%%%%%%%%%%%%%%%%%%%%%%%%%%%
%%%%%%%%%%%%%%%%%%%%%%%%%%%%%%%%%%%%%%%%%%%%%%%%%%%%%%%%%%%%%%%%%%%%%%%%%%%%%%%%%%%%

\end{document}